# Cellulose Photonic Pigments

Dr Richard M. Parker[1 †], Dr Tianheng H. Zhao[1 †], Dr Bruno Frka-Petesic[1], and Prof Silvia Vignolini[1 *]

[1] Yusuf Hamied Department of Chemistry, University of Cambridge, Lensfield Road, Cambridge, CB2 1EW, United Kingdom.

[†] Richard M. Parker and Tianheng H. Zhao contributed equally to this work.

* Correspondence to Prof Silvia Vignolini (sv319@cam.ac.uk).

## Abstract

When pursuing sustainable approaches to fabricate photonic structures, nature can be used as a source of inspiration for both the nanoarchitecture and the constituent materials. Although several biomaterials have been promised as suitable candidates for photonic materials and pigments, their fabrication processes have been limited to the small to medium-scale production of films. Here, by employing a substrate-free process, structurally coloured microparticles are produced *via* the confined self-assembly of a cholesteric cellulose nanocrystal (CNC) suspension within emulsified microdroplets. Upon drying, the droplets undergo multiple buckling events, which allow for greater contraction of the nanostructure than predicted for a spherical geometry. This buckling, combined with a solvent or thermal post-treatment, enables the production of dispersions of vibrant red, green, and blue cellulose photonic pigments. The hierarchical structure of these pigments enables the deposition of coatings with angular independent colour, offering a consistent visual appearance across a wide range of viewing angles.

## Introduction

Colour is an essential form of communication in our society, with the production and use of dyes and pigments dating back to prehistoric times. In the last few decades, the available colour palette has been enlarged and diversified enormously by the development of so-called *interference* or *effect pigments*.[1,2] While such photonic pigments are increasingly found in a wide variety of consumer products, spanning from automotive paints and security inks to textiles, cosmetics and food, they often rely on the use of energy-intensive methods, synthetic polymers[3] or inorganic materials (whose extraction raises ethical concerns[4]). When considering sustainable approaches to photonic pigments, nature can be used as a source of inspiration in terms of both the nanoarchitecture and the constituent materials.[5–7] Inspired by the hierarchical arrangement of cellulose found in structurally coloured plants,[8] here we report a substrate-free, emulsion-based methodology to produce cellulosic photonic pigments with vibrant colour spanning across the entire visible spectrum.

It is well known that colloidal suspensions of cellulose nanocrystals (CNCs) spontaneously self-organise above a certain volume fraction into a cholesteric (i.e., chiral nematic) colloidal liquid crystal, whereby the individual nanoparticles align locally along a helicoidal structure.[9,10] Moreover, evaporating a cholesteric CNC suspension can lead to photonic films that display vibrant, iridescent colour.[10–12] However, the self-assembly of CNCs confined within a spherical geometry, such as that of an emulsified droplet, has failed to produce microparticles with structural colour in the visible range, despite the attempts of several research groups.[13–19] The colour of CNC-based photonic materials is defined by the periodicity of the underlying helicoidal nanoarchitecture, referred to as the 'pitch', $p$. In general, the value of the pitch is affected by suspension-related parameters (*e.g.* pH or ionic strength) and, once the suspension is kinetically arrested,[10] by the geometry in which self-assembly takes place. We previously reported that for a hierarchical spherical geometry undergoing isotropic contraction, the relationship that regulates the pitch with the CNC volume fraction scales as $p \propto \phi^{-1/3}$, instead of $p \propto \phi^{-1}$, as expected for vertically-aligned domains in a drying film.[12,20] This reduced compression leads to microparticles with pitch values in the micron-range, correlating to reflection in the infrared regime rather than at visible wavelengths. In this study the inherent limitation of CNC self-assembly under spherical confinement is overcome by exploiting the interfacial buckling of kinetically arrested microdroplets of an aqueous CNC suspension. This emulsion-based methodology combines a careful optimisation of the aqueous CNC suspension with the controlled removal of water from the buckled CNC microparticles, resulting in vibrant red, green, and blue dispersions of cellulosic photonic pigments. Furthermore, we show that the hierarchical structure of these pigments enables the production of coatings with angular independent structural colour.

## Results

### *Fabrication of structurally coloured CNC microparticles*

In order to produce CNC microparticles with visible colour, an aqueous CNC formulation was first identified that yielded the smallest obtainable pitch independent of any geometric constraints. It was found that films cast from the anisotropic phase of a commercial CNC suspension (Univ. Maine, 7.0 wt.%), in the presence of additional salt ([NaCl]/[CNC] = 100 μmol $g^{-1}$), have an average pitch of 141 ± 9 nm, which results in reflection in the ultraviolet region (**Supplementary Figure 1**). As summarised in **Figure 1a**, the same aqueous CNC suspension was then emulsified in hexadecane *via* a flow-focusing microfluidic device to produce monodisperse water-in-oil microdroplets (Ø ≈ 160 μm). Upon controlled, slow drying beneath a hexadecane layer, the randomly oriented cholesteric CNC domains inside each microdroplet merged and reorganised to form a radially-aligned Frank-Pryce monodomain structure (**Supplementary Video 1** and **Supplementary Figure 12**).[13,20,21] Further water loss triggered the onset of kinetic arrest, with the periphery of the droplet arresting first due to a radial concentration gradient within the droplet that arose from the evaporative flux at the interface. The earlier arrest at the droplet interface contributes to the formation of a shell that then buckled due to the interplay between compressive capillary forces (that shrink both the droplet radius and surface area as the droplet dries) and the mechanical resistance of the solidifying cholesteric CNC shell to a reduction in its total surface area.[22–25] Significant buckling of the radially-aligned cholesteric shell enhanced the pitch compression within the arrested droplet, allowing the inherent limit of the spherical geometry to be overcome and resulting in microparticles that display visible colour.

The initial buckling resulted in red-coloured microparticles (Ø ≈ 100 µm), however additional compression of the pitch was required to produce the dispersions of green and blue CNC microparticles shown in **Figure 1b**. This was controllably achieved by applying a thermal or solvent post-treatment to further collapse and buckle the microparticles, as summarised in **Supplementary Figure 2** and discussed in detail later. The irregular and buckled surface of the microparticles can be visualised by dark-field microscopy (**Figure 1c**). Unlike concentrically-ordered photonic microspheres,[26–28] the majority of the reflected light is at oblique angles, providing a less directional optical response,[29] as evidenced by the microparticles appearing less vibrant in bright-field microscopy (**Supplementary Figure 3**). Micro-spectroscopy of individual CNC microparticles confirmed that the colour arises from a single reflection peak that is predominantly left-circularly polarised (**Figure 1d**), validating the underlying helicoidal architecture is the same as observed for photonic CNC films.

***Accessing additional pitch compression via desiccation***

The change in the visual appearance of the CNC microparticles arising from the additional compression upon treatment with a polar solvent is exemplified in **Figure 2**. The dried microparticles were first washed with *n*-hexane to remove residual non-volatile hexadecane and surfactant. In this non-polar solvent, the microparticles do not swell and a predominately red colour was reflected (**Figure 2a).** When the *n*-hexane was subsequently evaporated, the microparticles appeared white due to strong surface scattering at the air-particle interface, but without a noticeable volume change (**Figure 2b**). The CNC microparticles were then immersed in methanol, leading to a small degree of swelling and the reappearance of the red colour due to a reduced refractive index contrast at the particle-solvent interface (**Figure 2c**). However, upon evaporation of methanol, the microparticles experienced a significant volume reduction and consequently a pitch compression (**Figure 2d**), such that subsequent re-immersion in *n*-hexane revealed blue colouration (**Figure 2e)**. Less polar solvents can be used to induce smaller volume contractions that result in intermediate colours, with the green CNC microparticles shown in **Figure 1** obtained with isopropanol. Notably, while repeated washing with isopropanol does not induce any further colour change, exposing the isopropanol-washed microparticles to a more polar solvent (*e.g.*, methanol) results in a further irreversible blueshift (**Supplementary Figure 4**).

The effectiveness of the methanol treatment on different sized microparticles was tested. It was found that although microparticles of three sizes (Ø ≈ 25, 80, 124 μm) all demonstrated a similar blue appearance after methanol treatment, the circular polarisation of the optical response differed (**Supplementary Figure 5**). While the smallest microparticles reflected predominantly left-circularly polarised light, similar to conventional drop-cast CNC films,[30] the largest microparticles showed comparable intensity in both circular polarisations (**Figure 2f-j**). This trend in polarisation response can be attributed to the distortion of the helicoidal architecture within the buckled microparticles. First, the local shear of the cholesteric domains causes a distortion of the CNC helicoidal arrangement, which results in the reflected light having an elliptical polarisation that is still predominantly left-handed, *i.e.* that can be decomposed in a main left-circularly polarised (LCP) component as well as a small right-circularly polarised (RCP) component.[12,31] Second, at a larger scale the buckling of the interface results in highly tilted cholesteric regions, effectively acting as birefringent retardation plate. This enables the conversion of incident RCP light into LCP light, which can be reflected by deeper cholesteric regions within the microparticles, and then converted back into RCP light.[32,33] These effects are expected to be more prevalent in larger microparticles since they are both thicker and more buckled (**Supplementary Figure 6**).

Given that methanol and isopropanol are able to form hydrogen bonds with water and thus act as dehydrants (*e.g.* methanol is routinely used to fix proteins[34,35]), the volume contraction and corresponding pitch reduction induced by washing the 'dry' CNC microparticles with a polar solvent was attributed to the removal of residual water. This was verified by studying the effect of thermal treatment on the hexane-washed microparticles. Specifically, the microparticles were exposed to a constant temperature chosen between 40 and 200 °C for sixty minutes in a dry nitrogen atmosphere (**Supplementary Figure 7**), followed by optical microscopy in refractive index-matching oil ($n$ = 1.55) at ambient temperature (**Figure 3a**). Despite an apparent linear decrease in diameter with temperature, the CNC microparticles heated below 100 °C present only minor colour change suggesting this initial volume contraction had little impact on the buckled surface (**Supplementary Figure 8a**). However, samples exposed to temperatures above 100 °C displayed a more significant blueshift, with microparticles heated at 200 °C exhibiting a blue appearance comparable to their methanol-washed analogues (**Supplementary Figure 8b** and **Figure 1d**). Importantly, while the solvent-treated CNC microparticles disintegrate upon immersion in water, the microparticles heated at 200 °C maintain their integrity and ability to reflect visible light (with a moderate redshift), possibly as a result of the additional thermal desulfation of the CNCs (**Supplementary Figure 9**).[36–39]

To quantify the amount of residual water in the CNC microparticles, samples with and without methanol treatment were characterised by thermogravimetric analysis (TGA), as reported in **Figure 3b**. As expected, both curves exhibit a large mass loss starting at 250 °C, corresponding to the thermal degradation of cellulose.[40] However, prior to reaching this temperature, a significant mass loss (*ca*. 35%) was recorded in the sample without methanol treatment, while a much smaller mass loss (*ca*. 10%) was observed for the methanol-treated sample. After correcting for desulfation (see **Supplementary Discussion 1**), the mass fraction of removable water was calculated to be approx. 30% for the untreated sample. Scanning electron microscopy (SEM) measurement of the highly buckled CNC microparticles showed a decrease in the average diameter from 88.5 ± 0.8 µm to 77.9 ± 0.5 µm upon methanol treatment (**Figure 4a,b** and **Supplementary Table 1**). This corresponds to a 12.1 ± 1.1% diameter decrease, which, given it underestimates the impact of buckling, is in good agreement with the value of 13.9% estimated from the TGA analysis and calculated in **Supplementary Discussion 1**. Another important indication of the dehydrant effect of methanol is the reduced mass loss upon pyrolysis above 250 °C, which suggests that reaction with water to form CO and $CO_2$ no longer occurs (**Supplementary Figure 10**).[41] This further supports our hypothesis that both solvent and thermal treatment cause a blueshift in the reflected colour by the removal of residual water.

As expected from the optical analysis, SEM cross-sections of the microparticles revealed that the cholesteric domain remains locally well-aligned to the highly buckled surface **(Figure 4c,d**). The pitch measured in regions away from the hinge of a fold ($p_{\mathrm{limb}}$, which are primarily responsible for the visual appearance) are consistent across multiple regions and microparticles, with the average pitch $\langle p_{\mathrm{limb}} \rangle$ reduced from 396 ± 31 nm to 269 ± 16 nm after the methanol treatment, consistent with the observed optical response (**Figure 4e,f**). In contrast, the pitch measured in the hinge of a fold ($p_{\mathrm{hinge}}$, contributing to a negligible volume fraction) was much more variable and depended on the fold tightness, with $p_{\mathrm{hinge}} \approx \langle p_{\mathrm{limb}} \rangle / \cos\theta$, where the fold angle $\theta$ is defined in Figure 4. Interestingly, the percentage of pitch compression in the limbs is *ca*. 32%, which matches well with the volume contraction of the microparticle (32.1 ± 1.5%) rather than the reduction in diameter (12.1 ± 1.1%) as estimated from SEM (*cf.* Figure 4a,b). This indicates that buckling of the spherical microparticle allows the pitch to be locally compressed more significantly than expected for this geometry.[20] This greater compression, approaching that of a drop-cast film, is what enables reflection at visible wavelengths to be achieved. A multi-step buckling process, as suggested by the SEM images, can explain the mechanism of pitch compression and the route to volume shrinkage. Upon each buckling event during the microparticle drying process, $p_{\mathrm{hinge}}$ would stop contracting and thus would "record" a transient pitch, while $p_{\mathrm{limb}}$ would continue to shrink uniformly (with a corresponding $\theta$ increase), giving rise to coherent colour across the microparticle. This process would repeat, resulting in additional buckling events, leading to several generations of hinges where the oldest folds have the largest $p_{\mathrm{hinge}}$ and $\theta$. While buckling events facilitate the removal of water from the microparticles (as described below), they also increase the ability for the resulting distorted structure to withstand greater mechanical stress. In contrast to a drop-cast CNC film, where complete collapse of the cholesteric phase is achieved upon evaporation, this behaviour makes it increasingly difficult for residual water to escape upon drying under ambient conditions. As such the removal of this trapped water requires additional mechanical deformation, achievable either by displacement with a polar solvent or by thermal treatment, to allow capillary forces to further contract the microparticle.

***The role of interfacial buckling in achieving visible colour***

The mechanism leading to visible colour can be clarified by tracking the evolution of the cholesteric pitch within a droplet as it dries to form a buckled microparticle. The relationship between the pitch and the CNC volume fraction (determined from the diameter change) was measured by recording a time-lapse of a drying droplet using transmission polarised optical microscopy (**Figure 5a**: triangles). A clear transition is observed at 6.4% *v*/*v* (*i.e.* 9.9 wt.%), which is ascribed to kinetic arrest.[42] Beyond this point the pitch trajectory switches from $\phi^{-1.9}$ to $\phi^{-1/3}$, the latter being the scaling law expected for a completely arrested Frank-Pryce cholesteric structure in an isotropically shrinking sphere.[20] The pitch of the final untreated and methanol-treated microparticles, as measured in cross-section by SEM, is also plotted (Figure 5a: diamonds), with the corresponding CNC volume fractions determined from the TGA analysis. These values are clearly smaller than the projection of the $\phi^{-1/3}$ scaling law, suggesting that another mechanism intervenes. If this discrepancy can be attributed to buckling, the isoperimetric quotient, $Q = 4\pi\Sigma_2/\mathcal{P}_2^2$, defined from the cross-section area of the microparticles, $\Sigma_2$, and their corresponding cross-section perimeter, $\mathcal{P}_2$, can be used to quantify it (see **Supplementary Discussion 2**). From mass conservation arguments, $Q$ (which is smaller than unity for non-circular cross-sections) must be equal to the scaling factor impacting the particle contraction in the radial direction, and thus allows for an estimation of the radial pitch contraction due to buckling. A statistical analysis on both untreated and methanol-treated microparticles estimates $Q$ to be approximately 0.45 and 0.37, respectively (**Supplementary Figure 11**). This allows for an estimation of a hypothetical 'unbuckled pitch', which is found to follow the previously expected $\phi^{-1/3}$ scaling law (Figure 5a: circles), verifying this interpretation. As such, the anisotropic buckling of the drying CNC microdroplet allows the power law to evolve from that expected for isotropic spherical compression ($\phi^{-1/3}$) to one that approaches that of unilateral compression (*i.e.*, $\phi^{-0.86} \to \phi^{-1}$), as discussed further in **Supplementary Discussion 2**. This additional pitch contraction is key to accessing visible wavelengths, suggesting that for the production of photonic pigments an early kinetic arrest is vital to allow for the buckling to sufficiently develop during the drying process. By optimising both suspension parameters (*e.g.*, ionic strength) and the cellulose source, the CNC suspension employed in this study becomes kinetically arrested at a sufficiently low volume fraction (6.4% *v*/*v*). Combining this with the small initial pitch in suspension (facilitated by a steep $\phi^{-1.9}$ power law prior to kinetic arrest) and the use of only the fully anisotropic phase (expediting an early radial self-organisation upon emulsification), it allows for visibly coloured CNC microparticles to be achieved.[20] For completeness, a dish-cast film with vertically-aligned domains was also prepared from the same CNC suspension (**Supplementary Figure 1**), and the pitch was found to agree with the $\phi^{-1}$ scaling law expected for the film geometry when assuming the same kinetic arrest transition as for the drying microdroplets.[12,20]

Two different mechanisms can be independently responsible for the anisotropic compression that results in buckling of the kinetically arrested Frank-Pryce structure. First, an arrested cholesteric suspension may preferentially compress along the helical axis upon water loss. Indeed, strong alignment of rigid fibres can lead to anisotropic swelling or contraction that is favoured in the perpendicular direction,[43,44] and to anisotropic Young moduli in the dry state.[45] For the radial helical arrangement of the Frank-Pryce structure, this results in an imbalance of radial to orthoradial compression upon volume contraction,[46,47] leading to an excess of surface area to maintain a spherical shape. Second, the concentration front propagating inwards in the arrested droplet can lead to a more rigid shell that resists orthoradial compression upon further volume contraction.[48] Indeed, buckling of a spherical shell was reported for drying droplets of isotropic colloidal particle suspensions,[24] while it was prevented when high solvent permeability was maintained,[49] suggesting that buckling can be solely driven by the kinetics of water loss. In the latter case, an insufficient orthoradial contraction is compensated by an enhanced radial compression. Regardless of the precise mechanism, the high resistance of the drying microparticle to orthoradial compression prevents volume contraction if the spherical shape is maintained, so buckling is a necessary condition for further water removal. Interestingly, buckling also comes with high deformation costs and makes further volume contraction more difficult. As it contracts, the concentric cholesteric structure undergoes several buckling events resulting in several generations of wrinkles , which are increasingly able to resist further orthoradial compression despite the low water vapour pressure. This explains why capillary forces were not able to completely compress the 'dry' microparticles in hexane, while further desiccation with either polar solvent or by heat treatment allowed for further compression and buckling with higher order wrinkles.

An important consequence of anisotropic compression and buckling is a steeper pitch decrease than that predicted by the power law for an isotropic contraction ($p \propto \phi^{-1/3}$). To understand how the formulation of the initial CNC suspension influences the formation of a Frank-Pryce structure and its evolution into a buckled, coloured microparticle, three variations were prepared and compared against the standard (anisotropic) formulation (**Figure 6**), with the self-assembly also monitored over time (**Supplementary Figures 12-15**). A 50% dilution of the standard CNC suspension resulted in a more pronounced core-shell morphology within the drying droplets, resulting in irregular, collapsed microparticles with larger-period wrinkles. This morphology led to more inter- and intra-particle variations in colour, however it is notable that all microparticles were blue-green after subsequent methanol treatment. In a second variation, the isotropic fraction of the original CNC suspension was emulsified instead of the anisotropic fraction. The inability of this suspension to assemble into a Franck-Pryce structure within the microdroplet resulted in the formation of more spherical microparticles with only fine surface buckling. Faint structural colour arises from this thin shell, while the disordered, polydomain core leads to significant broadband scattering (i.e., whiteness). In a third variation, electrolytes (i.e., NaCl) were not added to the CNC suspension, resulting in a fully anisotropic and viscous suspension at 7.0 wt.%. Although the resulting microparticles were buckled, weak cyan colour was only apparent after methanol treatment. This redshift in colour was attributed to a larger initial pitch, with the untreated microparticles initially in the infrared, while the lower intensity is explained by the inability of this viscous suspension to fully develop into a well-ordered Frank-Pryce arrangement prior to becoming kinetically trapped. The origin of these various buckling morphologies and how they lead to such diversity in optical response is further discussed in **Supplementary Discussion 3**, with the self-assembly pathway summarised schematically in **Supplementary Figure 16**. Importantly, these results affirm that the anisotropy of the cholesteric contraction is important in the evolution of the buckled microparticles, particularly at low ionic strength, but can be further enhanced by kinetically trapping the droplet into a distinct core-shell morphology.

***Angular-independent colour from photonic pigments***

To demonstrate the potential use of the CNC microparticles as photonic pigments within a paint or coating, green microparticles were embedded in a polydimethylsiloxane (PDMS) film and optically characterised under two representative observation conditions. In the first configuration, the embedded film was observed at a fixed illumination angle ($\alpha_i$) with respect to the viewing direction and the sample itself was rotated at different angles ($\alpha_s$) defined with respect to the viewing direction (**Figure 7a-b**). Direct observations showed green reflection with no detectable colour change for the explored angles ($\alpha_s$). Angular-resolved optical spectroscopy confirmed the optical response remained centred at the same wavelength (approx. 490 nm, **Figure 7c**). In the second observation configuration, the sample was illuminated at normal incidence and the viewing direction was scanned, which corresponds to maintaining equal illumination ($\alpha_i$) and sample ($\alpha_s$) angles, such that $\alpha_i = \alpha_s \in [0°, 90°]$, **Figure 7d**). In this case, only a limited blueshift in the peak wavelength (*i.e.*, $\Delta\lambda \approx 25$ nm, $\Delta\lambda/\lambda$ < 5%) was detected (**Figure 7e**). These two complementary configurations, performed in off-specular conditions as defined with respect to the film interface (*i.e.*, $\alpha_i \neq 2\alpha_s$), demonstrate the non-iridescent character of the microparticles when embedded in the PDMS matrix, which sharply contrasts with the strong iridescence observed for dish-cast CNC films.[12]

The origin of the angular independence can be explained by a simple model based on ray-tracing analysis. As the macroscopic angular optical response of the buckled CNC microparticles matches surprisingly well with that of a radially-aligned cholesteric sphere (**Supplementary Figure 17**),[28,50] each embedded CNC microparticle was approximated by a microsphere with a concentric multi-layer architecture (**Supplementary Figure 18**). Due to Snell's law, the light reflected from the microparticle has a critical maximum angle that allows for transmission through the PDMS-air interface (**Supplementary Figure 19**). This limits the range of Bragg angles that contribute to the observed macroscopic optical response. From this model, the maximum expected spectral shift is limited to $\Delta\lambda/\lambda < 0.5\%$ in the first configuration and $\Delta\lambda/\lambda < 6.5\%$ in the second configuration (**Figure 4b,d**), validating the use of these microparticles as non-iridescent photonic pigments.

## Discussion

Starting from a commercially available CNC suspension, we successfully prepared cellulose photonic pigments that can reflect colour across the full visible spectrum, overcoming the large pitch inherent to spheroidal CNC microparticles. The key to additional pitch compression was interfacial buckling, which was achieved by an appropriate initial CNC source and formulation, as well as the controlled removal of residual water by either thermal or polar solvent post-treatment. The interfacial buckling distorted the helicoidal architecture of the microparticles, giving rise to reflection of both LCP and RCP light, which is rarely observed for CNC films. Importantly, cellulose photonic pigments embedded in a matrix display angular independent structural colour, in contrast to the iridescence of drop-cast photonic films. Finally, while recent progress has been made to upscale photonic CNC film production,[39] using an emulsion-based route allows for pigments to be directly produced in a single step without the need for a substrate. Furthermore, due to the good colour tolerance with respect to the microparticle size, this approach should be transposable from microfluidics to larger scale emulsion methods, such as membrane emulsification, which would enable them to be produced *via* a continuous fabrication process. As such, these cellulose photonic pigments offer a sustainable, biocompatible and scalable solution to the colourant industry, where there is a demand to transition away from synthetic polymers and unrenewable minerals to those derived from natural materials.

## Methods

### *Cellulose nanocrystal suspension*

The cellulose nanocrystal suspension was purchased from the Process Development Center of the University of Maine (batch no. 2015-FPL-077). The as-received suspension was provided pH-neutralised (*i.e.*, negatively-charged CNCs due to $–OSO_3^-$ groups, with $Na^+$ counter-ions), and was used as received, *i.e.*, no tip sonication or thermal treatment was applied. The measured concentration of the suspension was 10.7 wt.% (A&D, MX-50 moisture analyser). Consistent with previous benchmarking studies,[51] the zeta potential and Z-average size were respectively recorded to be -43.5 ± 1 mV and 121 ± 2 nm (Malvern Zetasizer Nano ZS; [CNC] = 0.25 wt.%, [NaCl] = 5 mM, passed through a 0.8 µm cellulose acetate syringe filter). A 6.0 wt.% dilution of this commercial suspension had a conductivity of 477 ± 6 µS $cm^{-1}$ (Mettler Toledo InLab 752-6MM) and a pH of 4.8 (Mettler Toledo InLab Micro Pro-ISM; measured in the presence of an excess of $K^+$ ions[52]). Conductometric titration of an acidified CNC suspension (0.5 wt.%, 20 mL) against sodium hydroxide (0.01 M) revealed $[-OSO_3^{(-)}]$ = 398 µmol $g^{-1}$ and $[-COO^{(-)}]$ = 9 µmol $g^{-1}$. The acidified suspension was prepared by ion exchange with sulfuric acid, followed by extensive dialysis against deionised water. An example TEM image of the elongated, splinter-like CNC nanoparticles is included in **Supplementary Figure 20**, along with a phase diagram for the as-received suspension.

The commercial suspension was diluted with Milli-Q water and aqueous sodium chloride solution (0.1 M) to obtain suspensions with 7.0 wt.% of CNC and a [NaCl]/[CNC] ratio of 85 – 100 µmol $g^{-1}$. The suspensions were then equilibrated for 7 days to allow for phase separation (e.g., for 100 µmol $g^{-1}$: 82% anisotropic phase, with $c_{iso}$ ≈ 6.5 wt.% and $c_{aniso}$ ≈ 7.1 wt.%), and the denser anisotropic phase was used as standard to prepare the emulsions. Alternative formulations were prepared following a similar procedure, as summarised in the flow chart in **Figure 6**.

### *Microdroplet generation*

Monodisperse microdroplets were generated within a hydrophobic, etched-glass microfluidic device containing a 105 µm wide flow-focusing junction and a channel depth of 100 µm (Dolomite, #3000158). The dispersed phase was the aqueous CNC suspension and the continuous phase comprised of hexadecane (Sigma-Aldrich) with 2.0 wt.% of Span 80 surfactant (Sigma-Aldrich). To generate monodisperse water-in-oil droplets (Ø ≈ 160 µm), the aqueous and oil phases were injected into the microfluidic device *via* two syringe pumps (Harvard Apparatus, PHD 2000) with respective flow rates of 800 and 7200 µL $h^{-1}$ (Nb. droplets with other sizes were achieved by tuning the flow rate ratio). To produce microparticles, 200 µL of CNC droplets were collected into a 9 cm polystyrene Petri dish filled with 15 mL of 0.5 wt.% of Span 80 in hexadecane and left to dry, with a typical drying time of 2 – 3 days. Notably, microdroplets dried much more slowly (*i.e.*, over several weeks) did not show any differences in the resultant microparticle morphology or optical response.

### *Solvent treatment of the CNC microparticles*

The dried microparticles were washed several times with *n*-hexane (Fisher) to replace the non-volatile hexadecane and to remove residual surfactant. Residual *n*-hexane was evaporated from the microparticles under a flow of nitrogen to yield a fine, free-flowing powder. To further blueshift the CNC microparticles, they were immersed in either isopropanol (Fisher) or methanol (Fisher) for two minutes and again blown dry to evaporate any residual solvent. The red, green and blue samples presented in this work were prepared from a 7.0 wt.% CNC suspension, with the ionic strength and solvent treatment varied to fine-tune the final colour of the microparticles, as summarised in **Supplementary Table 2**.

### *Thermal treatment of the CNC microparticles*

The *n*-hexane washed microparticles were dried on a coverslip and placed in a heating chamber (Linkam Scientific, THMS600). Dry nitrogen gas was flushed through the chamber while the sample was held at a specific temperature for 60 minutes. The initial heating rate was set to 20 °C $min^{-1}$. The decrease in microparticle size was observed to occur over the first 30 minutes of heating (**Supplementary Figure 7**).

***Preparation of microparticle-embedded films***

Suspending the CNC microparticles in a PDMS matrix allows for them to be well dispersed within a film geometry, while also removing interfacial scattering from the individual microparticles. PDMS and the cross-linker (Sylgard 184 elastomer kit, Dow Corning) were mixed in a 10:1 weight ratio, prior to blending with a powder of green CNC microparticles. Air bubbles were then removed from the mixture under reduced pressure until the dispersion was no longer cloudy. The mixture was poured onto a glass slide and allowed to cure at room temperature for approximately 24 hours. The microparticles did not change colour during the curing process.

***Preparation of a uniformly aligned dish-cast photonic CNC film***[53]

2 mL of a 50% dilution of the aforementioned stock CNC suspension (*i.e.*, the anisotropic phase of a 7.0 wt.% CNC suspension with [NaCl]/[CNC] ratio 100 µmol $g^{-1}$) was placed in a petri dish (Corning 430588, Ø = 35 mm, nontreated polystyrene) and inserted in the gap between a pair of NdFeB magnets (First4magnets, F390-N42, L40 × W40 × H30 $mm^3$) with vertically-aligned magnetisation (vertical gap = 24 mm, estimated field $\mu_0 H \approx 0.6$ T). The CNC suspension was left to slowly dry over *ca*. 4 days under an upturned glass beaker (3 L) at ambient conditions and the resultant transparent film characterised by UV-vis spectrometry (Cary 4000 spectrophotometer) and cross-sectional SEM analysis (see below).

***Optical Characterisation***

Prior to imaging, the microparticles were dispersed in refractive index-matching oil (Cargille Series A, $n_D^{25}$ = 1.5500) to reduce broadband scattering from the particle-oil interface. Note, no swelling or colour change of the microparticles was observed when immersed in this oil. *Reflection optical micrographs* were collected with a customised Zeiss Axio scope A1 microscope fitted with a CCD camera (Eye IDS, UI-3580LE-C-HQ, calibrated with a white diffuser) using a Halogen lamp (ZEISS, HAL100) as a light source in Koehler illumination. All bright-field and dark-field images were taken with a Zeiss EC Epiplan-Apochromat objective (20x, NA 0.6), with the microscope configured such that the numerical aperture (NA) in illumination was limited by the NA of the objective. The reflected light could also be filtered with a quarter wave plate and a linear polariser mounted at different orientations to distinguish between left- or right-handed circularly polarised light. To perform *micro-spectroscopy*, the microscope was coupled to a spectrometer (Avantes, AvaSpec-HS2048) using an optical fibre (Avantes, FC-UV200-2-SR, 200 µm core size) in confocal configuration. The reflectance spectra were normalised in dark-field in left-circular polarisation against a white diffuser (Labsphere SRS-99-010) coated with the same refractive index oil (**Supplementary Figure 21**). The *time-lapse series* were recorded on an Olympus IX-71 inverted microscope, using Olympus UPlanFLN (10x, NA 0.30) and LUCPlanFLN (40x, NA 0.60) objectives with an additional 1.6x magnifying lens, and imaged with a CMOS camera (Pixelink PL-D725CU-T). Where noted, transmission images were collected through crossed polarisers with an additional full-wave retardation (i.e., tint) plate (Olympus, U-TP530) to both enhance the overall brightness and contrast of the image, and to indicate the orientation of the cholesteric domains. *Photographs* of vials containing microparticles dispersed in ethyl cinnamate (Sigma-Aldrich, $n_D^{20}$ = 1.558) were recorded under diffuse illumination (*i.e.*, fluorescent ceiling light) with a Samsung Galaxy S9+ smartphone.

***Angular-resolved optical spectroscopy***

A laboratory-built goniometer was used to analyse the angular response of the CNC microparticles. A white light source (Schott KL1500 set to 3300 K) was used to illuminate as a collimated incident beam with a spot size of approximately 6 mm and at a fixed angle of 0°. The detector was mounted on an arm attached to a motorised rotation stage and coupled the light into an optical fibre (1000 µm) connected to a spectrometer (Avantes AvaSpec-HS2048XL). The sample was mounted on a rotation stage at the centre of the goniometer. The recorded light intensity was normalised with respect to the white Lambertian diffuser in air, while the exposure time was adjusted using an automatised high-dynamic-range (HDR) method.[53] For the microparticle-embedded film, the first measurement was collected while the sample was rotated between -90° to +75° with respect to the illumination source (setting the origin of the angles), with the detector fixed at -15° with respect to the illumination source. In the framework of the observer, this corresponds to an angular scan of the sample orientation $\alpha_s$ = [+90,-75] (as defined by the normal of the film surface, **Figure 7b**) at fixed illumination angle $\alpha_i$ = +15deg. The second measurement was collected with the sample fixed at 0° and by scanning the detector between 0° to +90°. In the framework of the observer, this corresponds to an angular scan of $\alpha_i$ = $\alpha_s$= [0,90] (**Figure 7d**). For the microparticle dispersion, the vial was fixed, and the detector was scanned between +45° and +185° (**Supplementary Figure 18a**). Angular-resolved photography was taken with a digital camera at a fixed position (Nikon, D3200, with a 68 mm macro extension tube), with either the sample or light source rotated.

***Thermogravimetric Analysis (TGA)***

CNC microparticles were evaluated using a TGA/DSC 2 instrument (Mettler Toledo) using an aluminium pan and with the following parameters: a nitrogen flow of 100 mL min$^{-1}$, heating rate of 1 °C min$^{-1}$, temperature range of 30 – 300 °C, sample mass of 5 – 10 mg.

***Scanning electron microscopy (SEM)***

Micrographs were collected with a Mira3 system (Tescan) operated at 5 kV and a working distance of 6 – 7 mm. The samples were mounted on aluminium stubs using conductive carbon tape and coated with a 10 nm thick layer of platinum with a sputter coater (Quorum, Q150T ES). To image the exterior of the microparticles by SEM, the microparticles were dispersed in *n*-hexane and the suspension was drop cast on to a glass coverslip followed by drying with nitrogen. To image the interior of microparticles, they were cryo-fractured using the following protocol: (i) the microparticles in *n*-hexane were drop cast on a coverslip, (ii) embedded in cellulose butyrate acetate resin (10 wt.% cellulose butyrate acetate dissolved in ethyl acetate), (iii) transferred into a dry nitrogen atmosphere, (iv) frozen in liquid nitrogen for 5 minutes, (v) mechanically crushed.

***Transmission electron microscopy (TEM)***

Micrographs were captured using a Talos F200X G2 microscope (Thermo Scientific, FEI) TEM operating at 200 kV and a CCD camera. Samples were prepared as follows: (i) The CNC suspension was diluted in ultrapure water in two successive steps, first to 0.1 wt.% then to 0.005 wt.%. (ii) A TEM grid (Agar Scientific S160-3 Carbon film 300 mesh Cu) was plasma treated, followed by deposition of a drop of the 0.005 wt.% CNC suspension. This was left to sit on the grid for 90 s before gently removing all residual liquid with filter paper (Whatman). (iii) The CNCs were stained with uranyl acetate solution for 60 s, with excess liquid again removed by blotting.

## Data Availability

All raw data relating to this publication is freely accessible from the University of Cambridge data repository (https://doi.org/10.17863/CAM.72840).

## References

1. Maile, F. J., Pfaff, G. & Reynders, P. Effect pigments - Past, present and future. *Prog. Org. Coatings* **54**, 150–163 (2005).

2. Pfaff, G. *Special Effect Pigments: Technical Basics and Applications*. (Vincentz Network, 2008).

3. Green, D. S., Jefferson, M., Boots, B. & Stone, L. All that glitters is litter? Ecological impacts of conventional versus biodegradable glitter in a freshwater habitat. *J. Hazard. Mater.* **402**, 124070 (2021).

4. ten Kate, A., Schipper, I., Kiezebrink, V. & Remmers, M. Beauty and a beast: Child labour in India for sparkling cars and cosmetics. *Sticht. Onderz. Multinatl. Ondernem.* (2016).

5. Burg, S. L. & Parnell, A. J. Self-assembling structural colour in nature. *J. Phys. Condens. Matter* **30**, 413001 (2018).

6. Wilts, B. D., Whitney, H. M., Glover, B. J., Steiner, U. & Vignolini, S. Natural helicoidal structures: Morphology, self-assembly and optical properties. *Mater. Today Proc.* **1**, 177–185 (2014).

7. Frka-Petesic, B. & Vignolini, S. So much more than paper. *Nat. Photonics 2019 136* **13**, 365–367 (2019).

8. Vignolini, S. *et al.* Pointillist structural color in Pollia fruit. *Proc. Natl. Acad. Sci. U. S. A.* **109**, 15712–15715 (2012).

9. Onsager, L. the Effects of Shape on the Interaction of Colloidal Particles. *Ann. N. Y. Acad. Sci.* **51**, 627–659 (1949).

10. Schütz, C. *et al.* From equilibrium liquid crystal formation and kinetic arrest to photonic bandgap films using suspensions of cellulose nanocrystals. *Crystals* **10**, 199 (2020).

11. Parker, R. M. *et al.* The Self-Assembly of Cellulose Nanocrystals: Hierarchical Design of Visual Appearance. *Adv. Mater.* **30**, 1704477 (2018).

12. Frka-Petesic, B., Kamita, G., Guidetti, G. & Vignolini, S. Angular optical response of cellulose nanocrystal films explained by the distortion of the arrested suspension upon drying. *Phys. Rev. Mater.* **3**, 045601 (2019).

13. Li, Y. *et al.* Colloidal cholesteric liquid crystal in spherical confinement. *Nat. Commun.* **7**, 12520 (2016).

14. Li, Y. *et al.* Periodic assembly of nanoparticle arrays in disclinations of cholesteric liquid crystals. *Proc. Natl. Acad. Sci. U. S. A.* **114**, 2137–2142 (2017).

15. Wang, P. X., Hamad, W. Y. & MacLachlan, M. J. Polymer and Mesoporous Silica Microspheres

with Chiral Nematic Order from Cellulose Nanocrystals. *Angew. Chemie Int. Ed.* **55**, 12460–12464 (2016).

16. Levin, D. *et al.* Green Templating of Ultraporous Cross-Linked Cellulose Nanocrystal Microparticles. *Chem. Mater.* **30**, 8040–8051 (2018).

17. Suzuki, T., Li, Y., Gevorkian, A. & Kumacheva, E. Compound droplets derived from a cholesteric suspension of cellulose nanocrystals. *Soft Matter* **14**, 9713–9719 (2018).

18. Liu, Y. *et al.* Assembly of cellulose nanocrystals in a levitating drop probed by time-resolved small angle X-ray scattering. *Nanoscale* **10**, 18113–18118 (2018).

19. Wang, C., Paineau, E., Remita, H. & Ghazzal, M. N. Cellulose Nanocrystals in Spherical Titania-Sol Microdroplet: From Dynamic Self-Assembly to Nanostructured TiO x/C Microsphere Synthesis. *Chem. Mater.* **33**, 6925–6933 (2021).

20. Parker, R. M. *et al.* Hierarchical Self-Assembly of Cellulose Nanocrystals in a Confined Geometry. *ACS Nano* **10**, 8443–8449 (2016).

21. Robinson, C., Ward, J. C. & Beevers, R. B. Liquid crystalline structure in polypeptide solutions. Part 2. *Discuss. Faraday Soc.* **25**, 29–42 (1958).

22. Tsapis, N. *et al.* Onset of buckling in drying droplets of colloidal suspensions. *Phys. Rev. Lett.* **94**, 018302 (2005).

23. Salmon, A. R. *et al.* Microcapsule Buckling Triggered by Compression-Induced Interfacial Phase Change. *Langmuir* **32**, 10987–10994 (2016).

24. Wang, J. *et al.* Magic number colloidal clusters as minimum free energy structures. *Nat. Commun.* **9**, 5259 (2018).

25. Datta, S. S., Shum, H. C. & Weitz, D. A. Controlled buckling and crumpling of nanoparticle-coated droplets. *Langmuir* **26**, 18612–18616 (2010).

26. Song, D. P., Zhao, T. H., Guidetti, G., Vignolini, S. & Parker, R. M. Hierarchical Photonic Pigments via the Confined Self-Assembly of Bottlebrush Block Copolymers. *ACS Nano* **13**, 1764–1771 (2019).

27. Vogel, N. *et al.* Color from hierarchy: Diverse optical properties of micron-sized spherical colloidal assemblies. *Proc. Natl. Acad. Sci. U. S. A.* **112**, 10845–10850 (2015).

28. Geng, Y., Noh, J. H., Drevensek-Olenik, I., Rupp, R. & Lagerwall, J. Elucidating the fine details of cholesteric liquid crystal shell reflection patterns. *Liq. Cryst.* **44**, 1948–1959 (2017).

29. Zhao, T. H. *et al.* Angular-Independent Photonic Pigments via the Controlled Micellization of Amphiphilic Bottlebrush Block Copolymers. *Adv. Mater.* **32**, 2002681 (2020).

30. Revol, J.-F., Godbout, D. L. & Gray, D. G. Solid self-assembled films of cellulose with chiral nematic order and optically variable properties. *J. Pulp Pap. Sci.* **24**, 146–149 (1998).

31. Juárez-Rivera, O. R., Mauricio-Sánchez, R. A., Järrendahl, K., Arwin, H. & Mendoza-Galván, A. Quantification of optical chirality in cellulose nanocrystal films prepared by shear-coating. *Appl. Sci.* **11**, 6191 (2021).

32. Song, M. H. *et al.* Effect of phase retardation on defect-mode lasing in polymeric cholesteric liquid crystals. *Adv. Mater.* **16**, 779–783 (2004).

33. Caveney, S. Cuticle reflectivity and optical activity in scarab beetles: the rôle of uric acid. *Proc. R. Soc. London. Ser. B. Biol. Sci.* **178**, 205–225 (1971).

34. Fricke, J. & Tillotson, T. Aerogels: Production, characterization, and applications. *Thin Solid Films* **297**, 212–223 (1997).

35. Kiernan, J. A. *Histological and Histochemical Methods: Theory and Practice*. (Scion, 2015).

36. Bardet, R., Roussel, F., Coindeau, S., Belgacem, N. & Bras, J. Engineered pigments based on iridescent cellulose nanocrystal films. *Carbohydr. Polym.* **122**, 367–375 (2015).

37. Beck, S. & Bouchard, J. Auto-Catalyzed acidic desulfation of cellulose nanocrystals. *Nord. Pulp Pap. Res. J.* **29**, 6–14 (2014).

38. Vanderfleet, O. M. *et al.* Insight into thermal stability of cellulose nanocrystals from new hydrolysis methods with acid blends. *Cellulose* **26**, 507–528 (2019).

39. Droguet, B. E. *et al.* Large-scale fabrication of structurally coloured cellulose nanocrystal films and effect pigments. *Nat. Mater.* **21**, 352–358 (2022).

40. Lin, N. & Dufresne, A. Surface chemistry, morphological analysis and properties of cellulose nanocrystals with gradiented sulfation degrees. *Nanoscale* **6**, 5384–5393 (2014).

41. Kim, D. Y., Nishiyama, Y., Wada, M. & Kuga, S. High-yield carbonization of cellulose by sulfuric acid impregnation. *Cellulose* **8**, 29–33 (2001).

42. Honorato-Rios, C. *et al.* Fractionation of cellulose nanocrystals: enhancing liquid crystal ordering without promoting gelation. *NPG Asia Mater.* **10**, 455–465 (2018).

43. Tardy, B. L. *et al.* Exploiting Supramolecular Interactions from Polymeric Colloids for Strong Anisotropic Adhesion between Solid Surfaces. *Adv. Mater.* **32**, 1906886 (2020).

44. Okajima, M. K. *et al.* Anisotropic swelling in hydrogels formed by cooperatively aligned megamolecules. *RSC Adv.* **5**, 86723–86729 (2015).

45. Miller, N. A., Li, Z., Xia, W. & Davis, C. S. Buckling Mechanics Modulus Measurement of Anisotropic Cellulose Nanocrystal Thin Films. *ACS Appl. Polym. Mater.* **4**, 3045–3053 (2022).

46. Warner, M., Terentjev, E. M., Meyer, R. B. & Mao, Y. Untwisting of a Cholesteric Elastomer by a Mechanical Field. *Phys. Rev. Lett.* **85**, 2320 (2000).

47. Nardinocchi, P., Pezzulla, M. & Teresi, L. Anisotropic swelling of thin gel sheets. *Soft Matter* **11**, 1492–1499 (2015).

48. Stimpson, T. C., Cathala, B., Moreau, C., Moran-Mirabal, J. M. & Cranston, E. D. Xyloglucan Structure Impacts the Mechanical Properties of Xyloglucan-Cellulose Nanocrystal Layered Films-A Buckling-Based Study. *Biomacromolecules* **21**, 3898–3908 (2020).

49. Al Harraq, A. & Bharti, B. Increasing aspect ratio of particles suppresses buckling in shells formed by drying suspensions. *Soft Matter* **16**, 9643–9647 (2020).

50. Noh, J., Liang, H. L., Drevensek-Olenik, I. & Lagerwall, J. P. F. Tuneable multicoloured patterns from photonic cross-communication between cholesteric liquid crystal droplets. *J. Mater. Chem. C* **2**, 806–810 (2014).

51. Reid, M. S., Villalobos, M. & Cranston, E. D. Benchmarking Cellulose Nanocrystals: From the Laboratory to Industrial Production. *Langmuir* **33**, 1583–1598 (2017).

52. Guidetti, G. *et al.* Co-Assembly of Cellulose Nanocrystals and Silk Fibroin into Photonic Cholesteric Films. *Adv. Sustain. Syst.* **5**, 2000272 (2021).

53. Frka-Petesic, B., Guidetti, G., Kamita, G. & Vignolini, S. Controlling the Photonic Properties of Cholesteric Cellulose Nanocrystal Films with Magnets. *Adv. Mater.* **29**, 1701469 (2017).

## Acknowledgements

This work is dedicated to the memory of Professor Chris Abell (Univ. Cambridge), not only for the considerable advances he brought to the field of droplet-based microfluidics, but also for the mentorship and generosity he showed to the authors during the completion of this research. The authors also thank Thomas Parton and Zihao Lu for characterisation of the drop-cast CNC film and Benjamin Droguet for useful discussions and characterisation of the CNC formulation. This work was supported by: the European Research Council ([ERC-2014-STG H2020 639088] and [ERC-2017-POC 790518] to S.V., R.P., T. Z., & B.F.-P.), the BBSRC ([David Phillips Fellowship BB/K014617/1] to S.V. and [BB/V00364X/1] to S.V. & R.P.), the EPSRC ([EP/R511675/1] to S.V. & R.P.), and the Winton Programme for the Physics of Sustainability (PhD scholarship to T.Z.).

## Author Contributions

Experiments were designed by R.P. and T.Z.; photonic microspheres were fabricated and characterised by R.P and T.Z.; and SEM was performed by T.Z. and analysed by T.Z. and B.F.-P.; angular-resolved spectroscopy was performed by B.F.-P with T.Z; TGA was collected by T.Z. and analysed by B.F.-P. with T.Z.; the buckling model was developed by B.F.-P. and R.P.; the manuscript was written by R.P. and T.Z. with contributions from B.F.-P. and S.V.

## Competing interests

The authors declare no competing interests.

## Figures

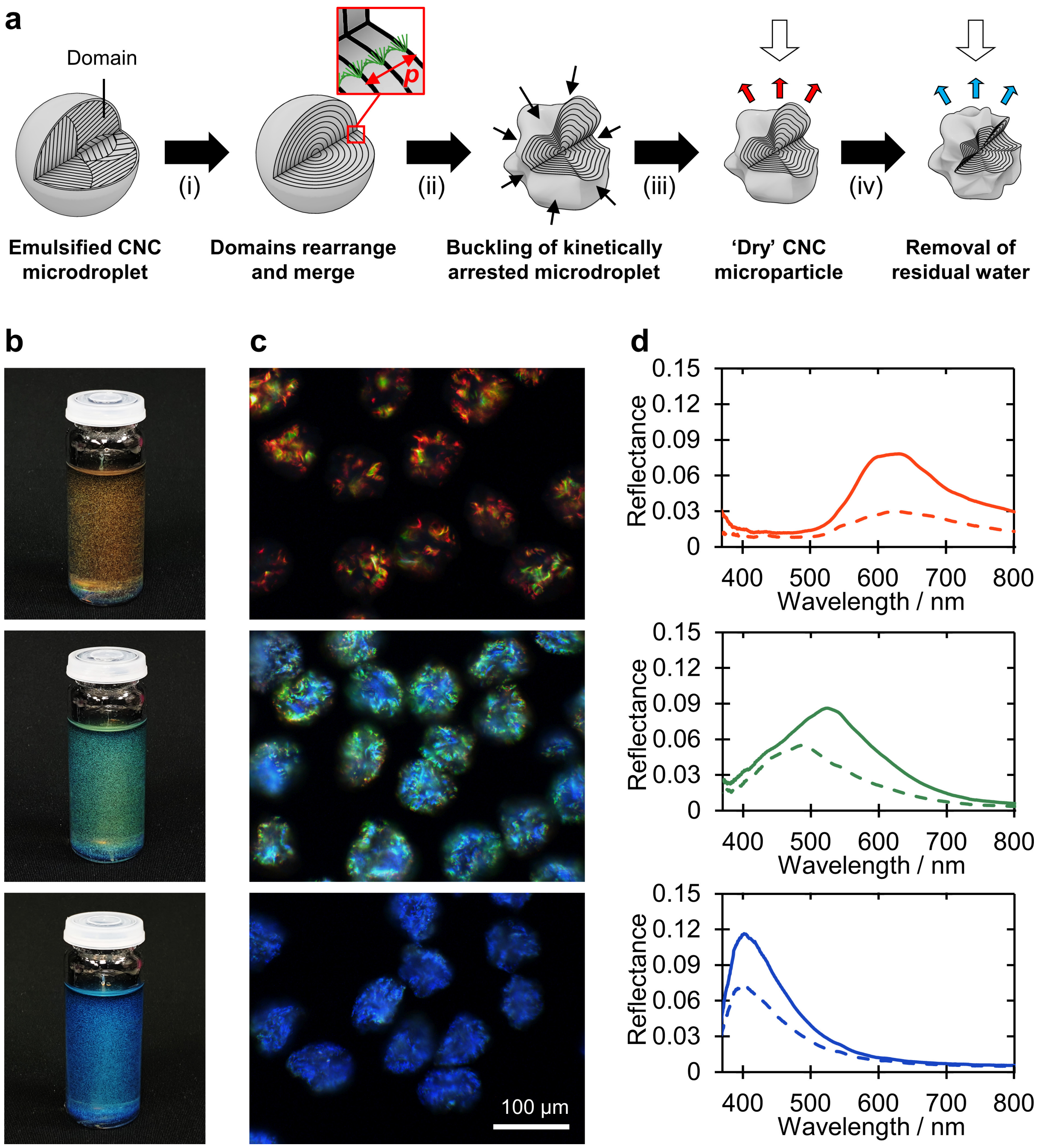

**Figure 1 | Photonic pigments *via* the confined self-assembly of cellulose nanocrystals. (a)** Schematic summarising the preparation of photonic microparticles: (i) the aqueous CNC suspension was emulsified in hexadecane *via* a microfluidic flow-focusing device; upon which the cholesteric domains self-organise and merge into a radially-aligned monodomain (Franck-Pryce structure); (ii) at a later stage of drying, the microdroplet becomes kinetically arrested and buckles; (iii) once drying is complete, CNC microparticles are formed with a complex surface morphology and visible red colouration; (iv) controlled removal of residual water within the microparticle induces additional buckling resulting in a further blueshift, enabling a full spectrum of photonic microparticles to be produced. **(b)** Photographs showing red, green, and blue cellulosic photonic pigments suspended in ethyl cinnamate (*ca*. 3.5 mg $mL^{-1}$, $n = 1.56$). The vial is 22 mm in diameter. **(c)** Dark-field microscopy images of individual CNC microparticles in refractive index oil ($n = 1.55$) and **(d)** associated micro-spectra of individual microparticles, collected through left-handed circularly polarisation (LCP, *solid line*) and right-handed circularly polarisation (RCP, *dashed line*) filters. The spectra are normalised against a white Lambertian diffuser coated with the same refractive index oil.

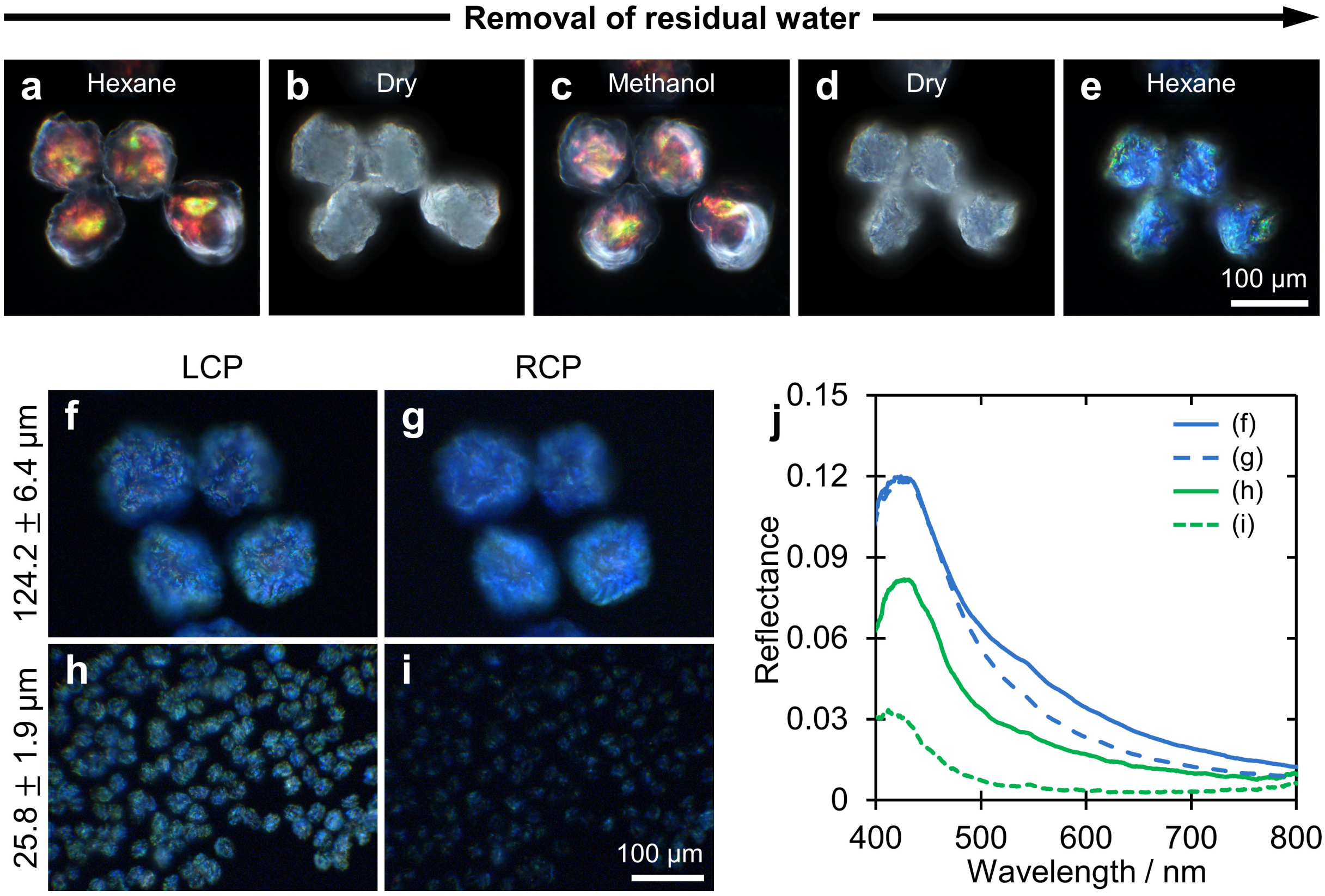

**Figure 2 | Optical analysis of CNC microparticles during solvent post-treatment. (a-e)** Sequence of dark-field micrographs showing the decrease in size and the blueshift in optical appearance after removal of residual water by methanol treatment. The loss of colour in the dry state arises from strong scattering at the microparticle-air interface. **(f-i)** Comparison of the optical appearance of large (*ca*. 124 µm) and small (*ca*. 25 µm) CNC microparticles in refractive index oil ($n$ = 1.55) when imaged through left-circular polarisation (LCP) and right-circular polarisation (RCP) filters; and **(j)** corresponding micro-spectra of individual microparticles, through LCP (*solid line*) and RCP (*dashed line*) filters and normalised against a white Lambertian diffuser coated with the same refractive index oil.

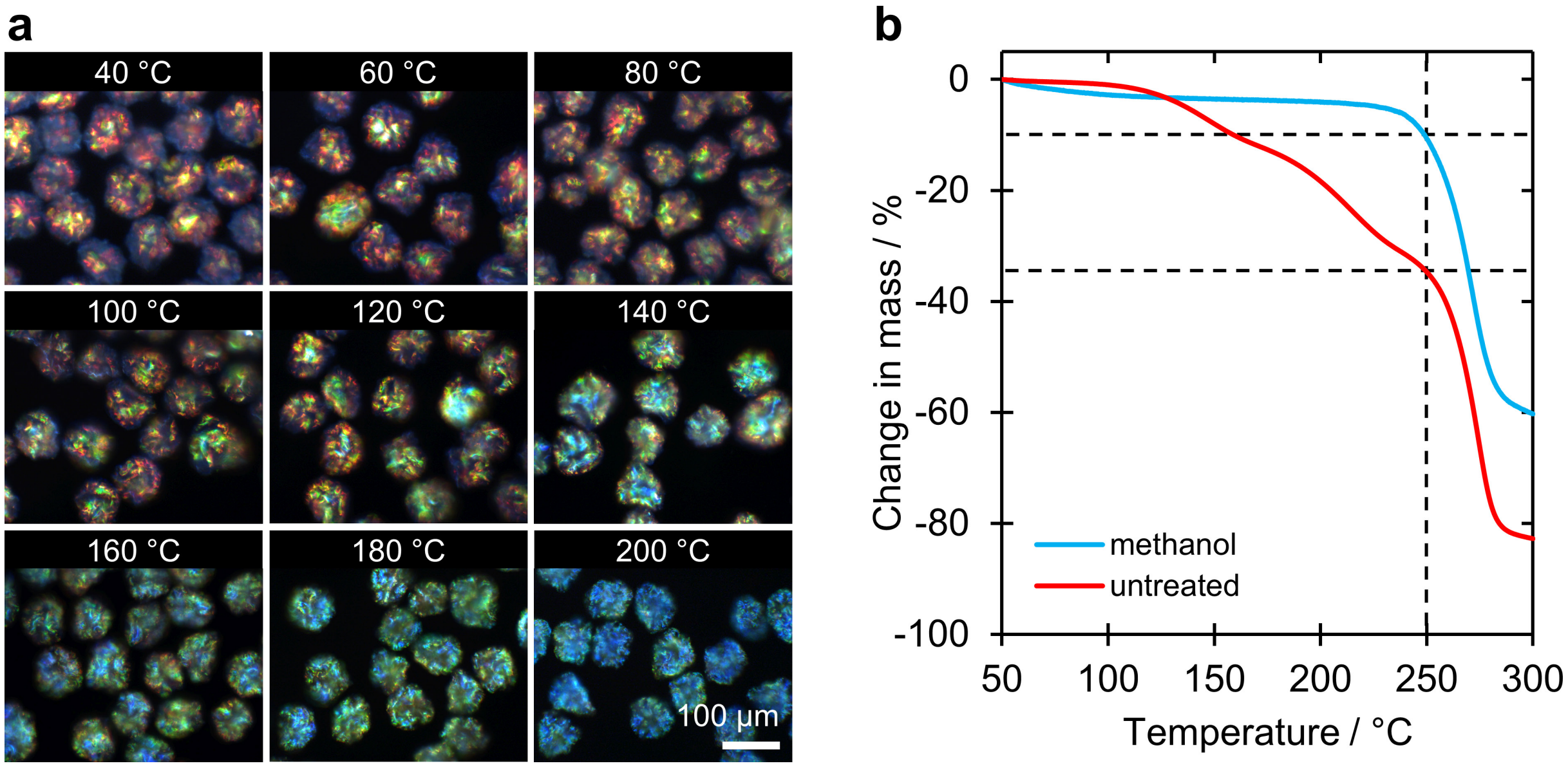

**Figure 3 | The impact of residual water on the optical appearance of CNC microparticles. (a)** Dark-field micrographs showing the effect of thermal treatment at different temperatures on the reflected colour of *n*-hexane washed CNC microparticles. **(b)** Thermogravimetric analysis (TGA) from 50 to 300 °C for a sample of CNC microparticles with, and without, treatment with methanol. The marked difference in the two curves is attributed to the removal of almost all the residual solvent trapped within the CNC microparticles.

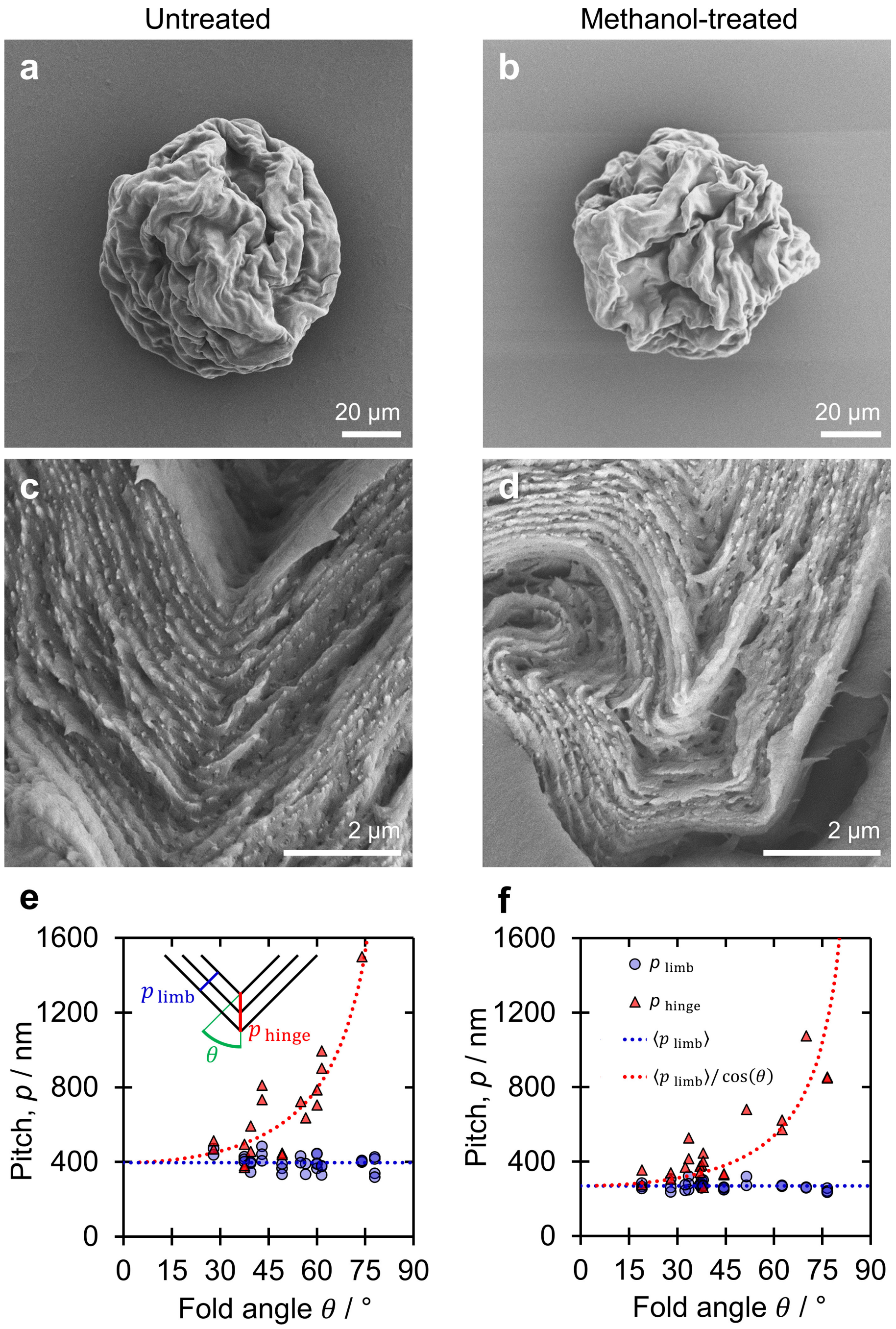

**Figure 4 | The role of interfacial buckling on the pitch of the radially-aligned CNC microparticles.** Scanning electron microscopy (SEM) images of a representative microparticle **(a)** prior to and **(b)** after treatment with methanol, showing the volume contraction and increase in buckling. Corresponding cross-sectional images in **(c)** and **(d)** show the related decrease in the pitch of the helicoidal structure. **(e,f)** Limb and hinge pitches of (e) hexane-washed and (f) methanol-treated microparticles plotted as a function of the fold angle $\theta$, as defined in the inset schematic. For both samples, the limb pitch is independent of $\theta$, while the hinge pitch generally follows the stated trigonometric relationship.

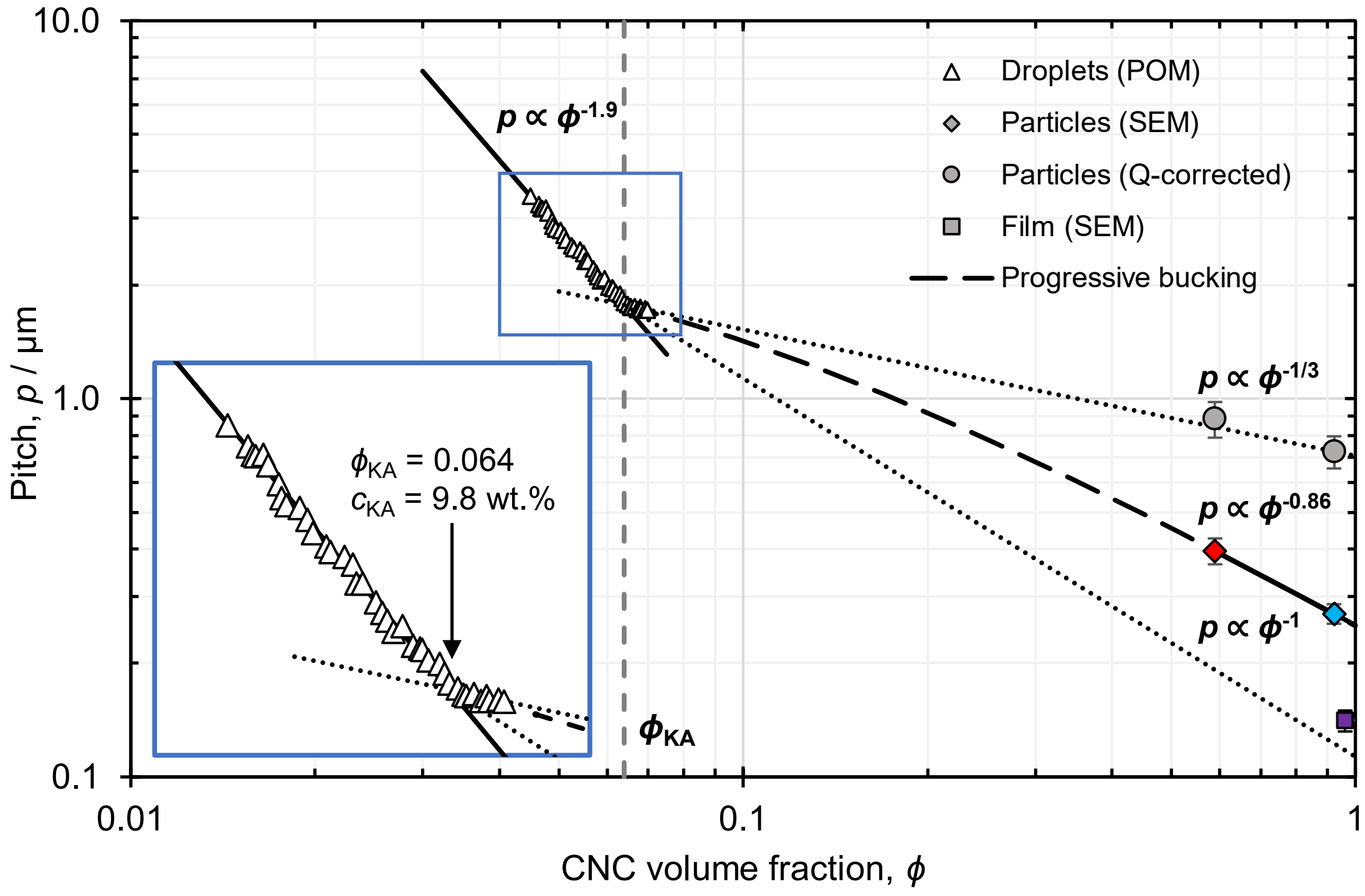

**Figure 5 | The reduction in the cholesteric pitch upon drying a microdroplet of CNC suspension to form a photonic microparticle.** The cholesteric pitch measured in the droplets by polarised optical microscopy (*triangles*) initially decreases following $p \propto \phi^{-1.9}$, until a transition at $\phi$ = 6.4% *v*/*v*, upon which it initially appears to follow $p \propto \phi^{-1/3}$ (see inset). An extrapolation of $p \propto \phi^{-1}$ from this transition is in reasonable agreement with SEM cross-sectional analysis of the corresponding film (*square*), validating this is the onset of kinetic arrest ($\phi_{\mathrm{KA}}$). Extrapolation of the $p \propto \phi^{-1/3}$, does not match with the pitch measured from the red and blue microparticles (*respectively red and blue diamonds*), with the additional pitch compression attributed to progressive buckling post kinetic arrest. By considering the isoperimetric quotient $Q$ of the microparticle cross-sections, the pitch they would have if buckling could be prevented can be estimated (*circles*), which is in good agreement with the $p \propto \phi^{-1/3}$ power law, confirming the key role of buckling in causing additional radial pitch contraction upon further compression. Error bars represent the average deviation of $p_{\mathrm{limb}}$ from SEM cross-sections.

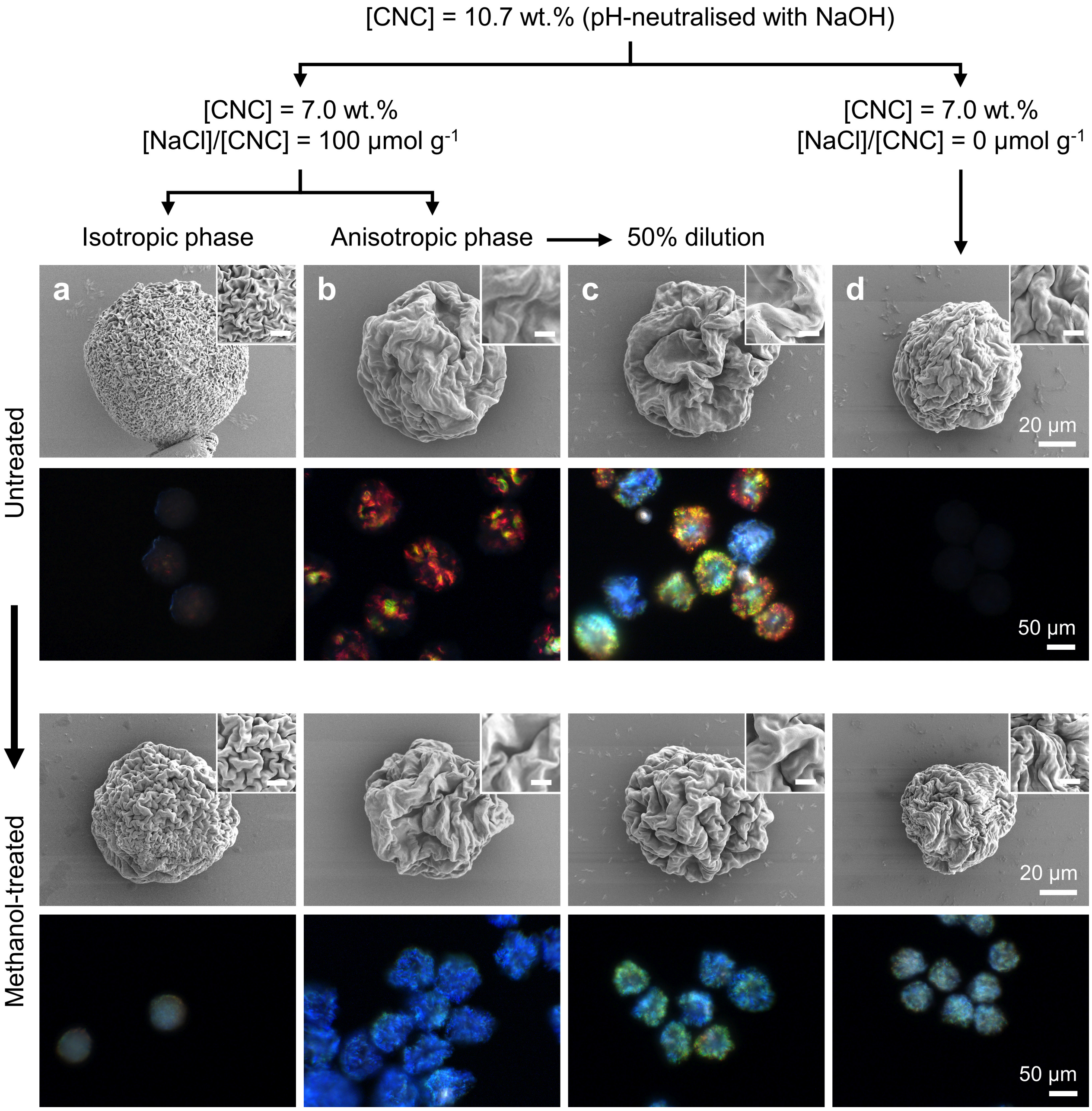

**Figure 6 | The role of CNC suspension formulation on the morphology and visual appearance of the photonic microparticles.** Scanning electron microscope and dark-field optical microscope images of CNC microparticles produced from different formulations of the commercial CNC suspension, and recorded prior to (*top*), and after methanol treatment (*bottom*): For the standard conditions employed in this study ([CNC] = 7.0 wt.% [NaCl]/[CNC] = 100 μmol $g^{-1}$), the suspension phase separates into **(a)** an isotropic phase and **(b)** an anisotropic phase. Microparticles produced from the anisotropic phase show vibrant colour, while those produced from the isotropic phase are only weakly coloured. **(c)** Diluting the anisotropic phase suspension in (b) by 50 vol% with water to yield an isotropic suspension prior to emulsification results in vibrantly microparticles with a range of colours, which can be blueshifted upon further solvent treatment. **(d)** Diluting the same CNC source in the absence of any additional salt ([NaCl]/[CNC] = 0 μmol $g^{-1}$) resulted in a fully anisotropic suspension that produced less intense, redshifted microparticles. The SEM inset scalebar is 5 μm.

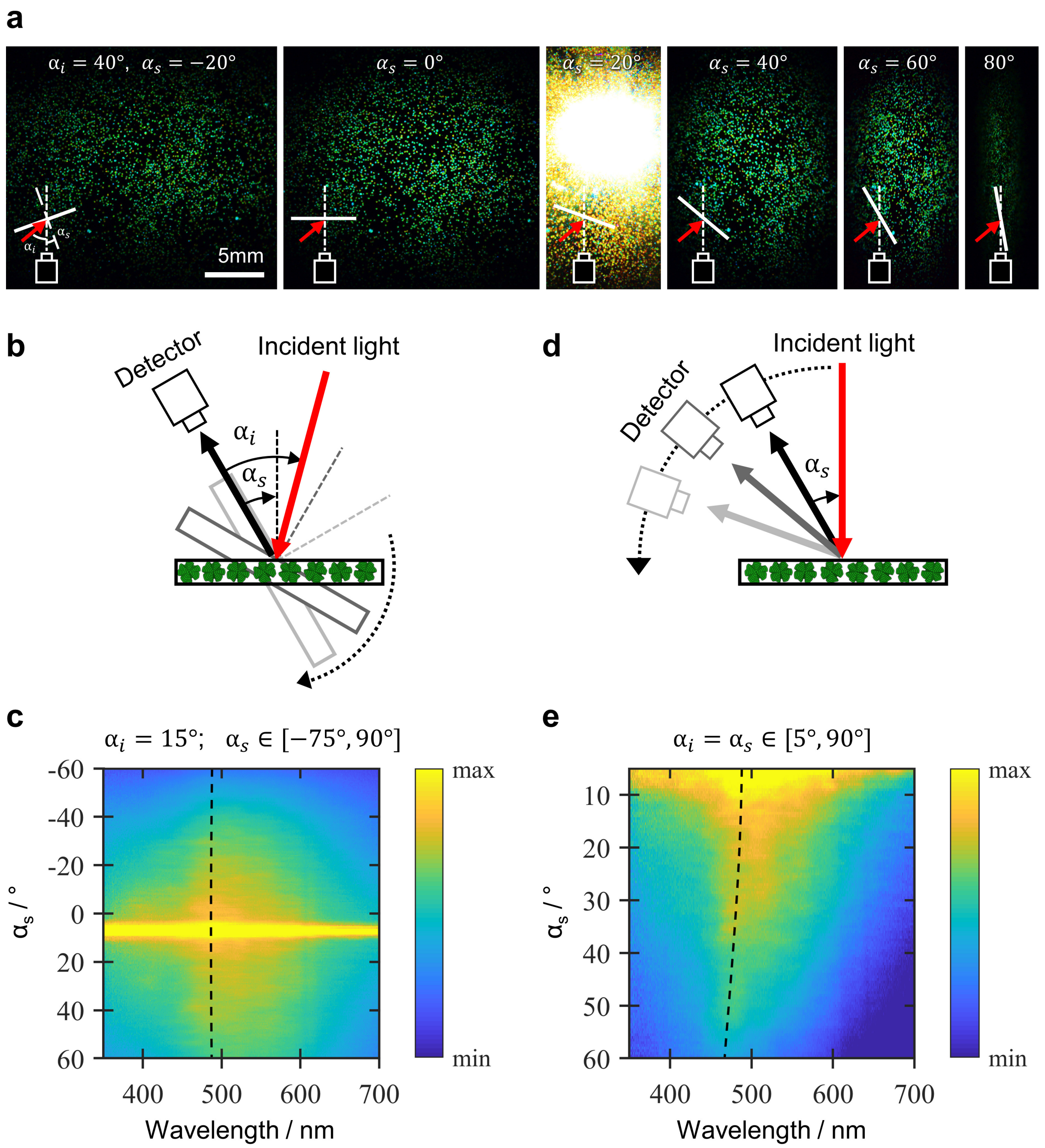

**Figure 7 | Angular dependence analysis of the reflected colour of coatings containing CNC photonic microparticles. (a)** Photographs showing a consistent green appearance for a PDMS film embedded with CNC photonic pigments, upon rotating the sample with a fixed illumination and viewing angle. Specular reflection from the film interface is at $\alpha_s$ = 20° **(b-e)** Angular resolved optical spectroscopy of the same PDMS film recorded under two complementary sets of illumination and viewing conditions, as described in the schematics. Both conditions show a limited angular dependence for the reflection peak over this wide angular range.

Supplementary Information

# Cellulose Photonic Pigments

Dr Richard M. Parker[1 †], Dr Tianheng H. Zhao[1 †], Dr Bruno Frka-Petesic[1], and Prof Silvia Vignolini[1 *]

[1] Yusuf Hamied Department of Chemistry, University of Cambridge, Lensfield Road, Cambridge, CB2 1EW, United Kingdom.

† Richard M. Parker and Tianheng H. Zhao contributed equally to this work.

* Correspondence to Prof Silvia Vignolini (sv319@cam.ac.uk).

# Supplementary Discussion

## Discussion 1: Analysis of TGA data

During the heating of both untreated and methanol-treated samples, the CNC microparticles experienced water loss, desulfation and eventually degradation of the cellulose, which all contribute to the mass loss observed in the TGA curves (Nb: TGA curves of the remaining mass fractions are subsequently referred to as $\text{TGA}_{\text{untreated}}(T)$ and $\text{TGA}_{\text{methanol}}(T)$ ). The methanol-treated sample experienced $w_{\text{MeOH},2} = \text{TGA}_{\text{methanol}}(50) - \text{TGA}_{\text{methanol}}(200) = 4\%$ of mass loss prior to 200 °C. This mass loss is likely to be due to residual methanol that was not fully removed after the solvent treatment. The expected volume decrease (with respect to 50 °C, which was taken as the reference) is thus expected to be $\frac{V_{200°\text{C},2}}{V_{50°\text{C},2}} = \frac{\rho_{\text{CNC}}^{-1} w_{\text{CNC},2}}{\rho_{\text{CNC}}^{-1} w_{\text{CNC},2} + \rho_{\text{MeOH}}^{-1} w_{\text{MeOH},2}}$ and $\frac{d_{200°\text{C},2}}{d_{50°\text{C},2}} = \left(\frac{V_{200°\text{C},2}}{V_{50°\text{C},2}}\right)^{1/3} \approx 0.9734$ using $\rho_{\text{CNC}}$=1.600 g cm$^{-3}$ and $\rho_{\text{MeOH}}$=0.792 g cm$^{-3}$ (and $w_{\text{CNC},2}$ = 1- $w_{\text{MeOH},2}$). From 200 °C to 250 °C, a further mass loss $w_{\text{sulf},2} = \text{TGA}_{\text{methanol}}(200) - \text{TGA}_{\text{methanol}}(250)$= 6% is attributed to the desulfation of the CNCs, so that $w_{\text{desulfCNC},2}$=$\text{TGA}_{\text{methanol}}(250) = 90\%$. If we assume this desulfation to be independent of the solvent, we can consider the relative mass loss due to desulfation in methanol-treated particles is proportional to the amount of remaining cellulose in the untreated particles, so $\frac{w_{\text{CNC},2}}{w_{\text{desulfCNC},2}} = \frac{w_{\text{CNC},1}}{w_{\text{desulfCNC},1}}$, where the subscripts 1 and 2 refer, respectively, to untreated and the methanol treated samples. The desulfated CNC mass fraction in the untreated sample is taken as $w_{\text{desulfCNC},1} = \text{TGA}_{\text{untreated}}(250)$, from which we get $w_{\text{CNC},1} = 69.65\%$ and $w_{\text{water},1} = 1 - w_{\text{CNC},1}$. The corresponding initial volume before water evaporation is estimated as $\frac{V_{50°\text{C},1}}{V_{200°\text{C},1}} = \frac{\rho_{\text{CNC}}^{-1} w_{\text{CNC},1} + \rho_{\text{water}}^{-1} w_{\text{water},1}}{\rho_{\text{CNC}}^{-1} w_{\text{CNC},1}} \approx 1.697$ and the diameter ratio as $\frac{d_{50°\text{C},1}}{d_{200°\text{C},1}} = \left(\frac{V_{50°\text{C},1}}{V_{200°\text{C},1}}\right)^{1/3} = 1.1928$. The diameter ratio expected between the untreated and treated particles is then estimated as $\frac{d_{50°\text{C},2}}{d_{50°\text{C},1}} \approx \frac{1}{1.1928} * \frac{1}{0.9734} \approx 0.8612$, i.e., a decrease of 13.9%.

Another relevant observation is the difference at which the two TGA curves seem to plateau at high temperature, indicating the mass of solid carbon atoms (C) remaining from the pyrolysis of CNCs: $w_{\text{C},1} = \text{TGA}_{\text{untreated}}(300) \approx 17.26\%$ and $w_{\text{C},2} = \text{TGA}_{\text{methanol}}(300) \approx 39.73\%$. After rescaling by their respective proportion of CNC content in each sample, we obtain the ratios $r_{(\text{C/CNC}),1} = \frac{w_{\text{C},1}}{w_{\text{CNC},1}} \approx 23.8\%$ and $r_{(\text{C/CNC}),2} = w_{\text{C},2}/w_{\text{CNC},2} \approx 39.7\%$. This has to be compared to the expected mass ratio of six carbon atoms in cellulose $(C_6H_{10}O_5)_n$, which gives $r_{(\text{C/CNC}),\max} = 44.45\%$ and represents a theoretical limit if all C atoms are retained in the solid residue, namely $\frac{r_{(\text{C/CNC}),1}}{r_{(\text{C/CNC}),\max}} = 53.5\%$ and $\frac{r_{(\text{C/CNC}),2}}{r_{(\text{C/CNC}),\max}} = 89.4\%$. According to D.-Y. Kim *et al.*,[1] the presence of adsorbed water allows for the reaction of oxygen atoms (present in water) to react with carbon and release CO and $CO_2$ gas, thus decreasing the ratio of the remaining solid carbon residue. This rescaling makes the excess water in the initial untreated particles more apparent, but also indicates that at 300 °C, methanol treatment significantly prevents mass loss *via* pyrolysis, approaching the theoretical maximum (green dashed line in Supplementary Figure 10). This confirms the role of methanol as a dehydrant, but also indicates that, since some CO and $CO_2$ has formed (both $r_{(\text{C/CNC}),1}$ and $r_{(\text{C/CNC}),2}$ < 1), the above estimations assuming complete dehydration at 250 °C are only indicative.

## Discussion 2: Scaling analysis of pitch compression in the presence of buckling

In the following discussion a qualitative description of the pitch scaling behaviour is proposed based on a stepwise introduction of shrinkage, local anisotropic distortion, and global buckling transformations, which are treated sequentially for the sake of clarity, but expected to occur in parallel. This pitch scaling is relevant only after the kinetic arrest of the suspension has occurred.

Qualitatively, the pitch shrinkage upon water loss that is expected in the absence of buckling for a kinetically arrested Frank-Pryce structure can be estimated from a simple scaling law. Let us define an initial spherical droplet of radius $R_0$, surface $S_0 = 4\pi R_0^2$, volume $V_0 = 4\pi R_0^3/3$ and pitch $p_0 = R_0/N$, where $N$ is a constant. The cross-sectional area of the droplet is $\Sigma_0 = \pi R_0^2$ and its perimeter is $\mathcal{P}_0 = 2\pi R_0$. The volume fraction of CNCs in the droplet is defined as $\phi_0$ and the solid angle of the surface viewed from the centre is $\Omega_0 = S_0/R_0^2 = 4\pi$. This initial state corresponds to the droplet at the onset of the kinetic arrest.

After a radial contraction $\alpha$ $(0 < \alpha \leq 1)$ in the absence of buckling, these quantities are expected to scale as $R_1 = \alpha\, R_0$, $S_1 = \alpha^2\, S_0$, $V_1 = \alpha^3\, V_0$, $p_1 = \alpha\, p_0$, $\Sigma_1 = \alpha^2 \pi R_0^2$ and $\mathcal{P}_1 = 2\pi\, \alpha R_0$, respectively. The volume fraction of CNCs in the droplet is given by $\phi_1 = \alpha^{-3}\phi_0$ and the solid angle of the surface viewed from the centre is unchanged $\Omega_1 = S_1/R_1^2 = 4\pi$.

Let us introduce a distortion term $\beta$ $(\alpha \leq \beta \leq 1)$ to account for a local asymmetry of the contraction of the cholesteric structure, whereby the compression along the helical axis is facilitated. The aforementioned quantities in this distorted particle can, in first approximation, be expected to scale as $R_2 = \beta^2\, R_1 = \alpha\beta^2 R_0$, $S_2 = \beta^{-2} S_1 = \alpha^2\beta^{-2}\, S_0$, $V_2 = V_1 = \alpha^3\, V_0$, $p_2 = \beta^2\, p_1 = \alpha\beta^2\, p_0$, $\Sigma_2 = \alpha^2\pi R_0^2$ and $\mathcal{P}_2 = 2\pi\, \alpha\beta^{-1} R_0$, respectively. The volume fraction remains unaffected, i.e., equal to $\phi_2 = \phi_1 = \alpha^{-3}\phi_0$, resulting in the relationship $\alpha = (\phi_2/\phi_0)^{-1/3}$ that will prove useful later. Note, the resulting solid angle of this larger surface, when viewed from the centre, is $\Omega_2 = S_2/R_2^2 = 4\pi\beta^{-6} \geq 4\pi$, which causes an expanded solid angle that is inconsistent to maintain an unbuckled spherical shape if $\beta < 1$. The distortion term $\beta$ intervenes thus both in the pitch $p_2$ and in the perimeter $\mathcal{P}_2$. We can estimate $\beta$ from cross-sectional SEM analysis of a buckled microparticle, by measuring $\Sigma_2$ and $\mathcal{P}_2$ and deducing from it $\beta^{-2} = \mathcal{P}_2^2/(4\pi\Sigma_2)$, as reported in Supplementary Figure 11. The inverse quantity, $Q = 4\pi\Sigma_2/\mathcal{P}_2^2$, is also known as the isoperimetric quotient, such that we get $Q = \beta^2 = p_2/p_1$. The pitch $p_1$ (a so-called '$Q$-corrected pitch' that represents the pitch in the absence of buckling) can thus be experimentally estimated from the measured $p_2$ of the microparticles and from their corresponding $Q$ using $p_1 = p_2/Q$.

While this modelling allows a prediction of the pitch $p_2$ for a given distortion term $Q$, it poses no constraints on the possible relationship between $Q$ and $\phi_2$, or in other words, between $\alpha$ and $\beta$. Various scenarios are possible for the pitch function $p_2(\phi)$, then defined as the pitch at a given volume fraction $\phi$ in the buckled droplet or particle rather than only in the final solid particle,. Several cases are worth discussing and are also illustrated in Supplementary Figure 22:

- The case $\beta = \alpha$ leads to a power law $p_2(\phi) = \alpha\beta^2\, p_0 = \alpha^3\, p_0 = p_0(\phi/\phi_0)^{-1}$ corresponding to vertical domains in a film geometry. This scaling is more complex for misaligned domains, which are less compressed resulting in structures with larger pitch values.[2,3] This is the reason why the dish-cast film prepared in this work was left to self-assemble under a vertical magnetic field, which favours vertical aligned domains.
- The case $\beta = 1$ leads to the unbuckled scenario $p_2(\phi) = \alpha\, p_0 = p_0(\phi/\phi_0)^{-1/3}$.
- The simple power law case, whereby buckling and kinetic arrest both occur at the same time and where the buckling term obeys a simple power law $\beta = \alpha^d$, is inconsistent with the pitch values observed in the hexane- and methanol-treated particles but agrees within the error bars with the $\beta$ values deduced from the isoperimetric quotient $Q$ for both treatments. If such a power law is forced through $(\phi_0, p_0)$ and $(\phi_2, p_2)$ of the blue particles, we obtain $\beta = \alpha^{0.55}$ and $p_2(\phi) = \alpha\beta^2\, p_0 = p_0(\alpha/\alpha_3)^{2.4} = p_3(\phi/\phi_3)^{-0.7}$.

- The delayed buckling simple power law case corresponds to:
  - a first stage without buckling at $\beta = 1$ for the values $\phi_0 < \phi < \phi_3 \approx 0.141$ (i.e., $1 > \alpha > \alpha_3 \approx 0.768$ ), leading to $p_2(\phi < \phi_3) = \alpha\, p_0$ and the pitch at the onset of buckling $p_3 = \alpha_3\, p_0$.
  - followed by a buckling regime at $\phi > \phi_3 \approx 0.141$ (i.e., $\alpha < \alpha_3 \approx 0.768$) following a simple power law, where $\beta = (\alpha/\alpha_3)^{0.794} = 1.233\,\alpha^{0.794}$ . This leads to $p_2(\phi > \phi_3) = \alpha\beta^2\, p_0 = p_3(\alpha/\alpha_3)^{2.588} = p_3(\phi/\phi_3)^{-0.863}$, matching the experimentally derived pitch power law between the hexane- and the methanol-treated microparticles.

  These allow for an estimation of a lower and an upper bound value for the buckling onset to $\phi_0 < \phi_{\text{Buck}} < \phi_3$ in the general case since a steeper behaviour than $\sim p_3(\phi/\phi_3)^{-0.863}$ for the pitch beyond $\phi_3$ is unlikely.

- The progressive buckling case is an *ad hoc* model describing a possible scenario, whereby both kinetic arrest and buckling onset occur at the same time, yet where buckling develops progressively as the contraction evolves to more extreme values. The model proposed here is just one among many possible and was constructed with $\beta \approx \alpha^{0.853-0.718\alpha}$ and $p_2(\phi) = \alpha\beta^2\, p_0 \approx p_0\alpha^{2.707-1.437\alpha} = p_0(\phi/\phi_0)^{0.479(\phi_0/\phi)^{1/3}-0.902}$ . While this model is purely phenomenological and may appear mathematically cumbersome, it illustrates well the possibility of more complex behaviours, whereby the buckling magnitude is initially weak, accounting for the relative compressibility of the soft cholesteric shell and the cholesteric domains perpendicular to their helical axes, and at a later stage gets much more pronounced, as the CNCs struggle to pack efficiently in their alignment direction, compared to the perpendicular direction.

## Discussion 3: Role of the CNC formulation on the microparticle morphology

***Investigating the effect of CNC formulation on the confined self-assembly pathway***

To understand how the initial formulation of the CNC suspension impacts the formation of a Frank-Pryce cholesteric structure and its evolution into a buckled microparticle, three variations were prepared (see Figure 6 of the main article), with their self-assembly pathways also monitored over time and compared to the standard suspension (Supplementary Figures 12-15). From this, a proposal of the self-assembly route for each formulation variation was derived, as depicted schematically in Supplementary Figure 16.

Firstly, it was found that a simple two-fold dilution of the standard CNC suspension yielded microparticles with a much larger initial variation in colour (Figure 6c). However, upon treatment with methanol, all of the microparticles shifted to blue wavelengths. This dilution likely favoured a greater concentration gradient upon drying within the initially isotropic droplet, promoting the formation of a kinetically arrested Frank-Pryce shell, surrounding an isotropic core (Supplementary Figure 13). Upon further drying, the distinct core region continues to concentrate until it becomes anisotropic, but does not adopt radial alignment, likely due to being disfavoured by the small volume and distortion of the spherical geometry upon initial buckling. This gives rise to various shapes and water content, associated with different abilities to resist compression against centripetal capillary forces. The significant uniaxial (i.e., radial) collapse of the shell during the initial buckling process, when the liquid core is still compliant, is probably at the origin of the resulting intra- and inter-particle diversity in colour. In contrast, the more uniform blue-green colour observed after methanol treatment can then be explained by the overpowering dehydration effect that occurs regardless of the remaining water content, leading eventually to a more comparable pitch compression for all particles and orientations upon reswelling and subsequent collapse. These results suggest that the existence of an asymmetry of the mechanical properties between a softer core and a more rigid shell favours buckling, but that a large difference is not desirable.

Secondly, microparticles produced from the corresponding isotropic phase of the standard phase-separated CNC suspension were much more spherical, with only fine surface buckling observable under SEM (Figure 6a). It was found that such microparticles display only very weak structural colour, with significant broadband scattering. Furthermore, while methanol treatment again resulted in a contraction in diameter and a blueshift, it did not enhance the intensity. These observations could be explained by the time-lapse series, which revealed that while tactoids do form upon drying the initially isotropic droplet, they do not have sufficient time to merge and organise, resulting in the formation of only a thin Frank-Pryce shell around a mostly polydomain droplet (Supplementary Figure 14). This arrangement explains the presence of only fine surface buckling and supports the idea of a thin rigid shell (with smaller pitch, where any structural colour originates) surrounding a polydomain core that did not buckle (with larger pitch, which simply scatters visible light).

Thirdly, microparticles produced from a CNC suspension formulated without any additional salt appeared colourless, which is attributed to the particles reflecting at near infrared wavelengths (Figure 6d). Such a redshift is expected upon reducing the ionic strength of the initial CNC suspension, due to the greater double layer repulsion between the CNC nanorods that typically screen the chiral interactions between CNCs, resulting in an increased pitch. Upon methanol treatment, the microparticles were observed to collapse and blueshift, but this initial offset remained, resulting in cyan rather than blue colouration. The time-lapse series showed that while the anisotropic droplet can relax and self-organise to a degree, the much higher viscosity of this suspension inhibits its ability to fully develop into a Frank-Pryce arrangement, prior to becoming trapped presumably as a Wigner glass (Supplementary Figure 15). As such, this disordered Frank-Pryce has consequently much weaker reflectivity than for the standard case (Figure 6b), confirming that long-range order is necessary to maximise the optical response. In the absence of added ions, CNCs are expected to repel each other more strongly, causing a more even distribution of the CNCs across the droplet. Unlike the previous two cases, the presence of comparable structural colour supports the idea that a strong mechanical asymmetry between core and shell is not essential and that the anisotropic compressibility of the cholesteric phase parallel vs perpendicular to the helical axis (i.e., radial vs orthoradial for a Frank-Pryce structure) could be sufficient to make the desired pitch decrease upon drying.

***Comparison with microparticles prepared from a laboratory-made CNC suspension***

In light of this new understanding of the role of buckling and residual water on the pitch of CNC microparticles, it is interesting to compare these observations with our previous article,[4] cited as reference 20 in the main article, where the microparticles were significantly less buckled and the measured pitch was in the micron range. Among the important differences to note, we highlight the following:

- The previous CNC suspension was laboratory-made from cotton, it was tip-sonicated, then heat-treated and no extra electrolytes were added to the formulation. The produced microparticles were not exposed to a desiccating post-treatment such as solvent or heat.
- The CNC suspension in the present work is commercially made from wood pulp and has never been sonicated. It was diluted to 7 wt.% with added NaCl to form a biphasic suspension, from which the lower anisotropic phase was used to make the emulsion. The colour of the resultant microparticles was tuned with either polar solvent or thermal post-treatments, to achieve additional desiccation.

As the two protocols differ in several aspects, identifying the key factors leading to pitch reduction and thus structural colour in the visible regime is not straightforward. For example, tip sonication is known to increase the pitch in suspension and redshift the photonic response in dried films, while heat treatment is known to cause desulfation, resulting in reduced surface charge and release of electrolytes, both leading to a blueshift. However, the impact of such treatments on the onset of kinetic arrest, is still not well understood. This is particularly important in the context of emulsified droplets where kinetic arrest leads to both significant changes in the scaling power law for pitch contraction and potentially the onset and extent of buckling.

For this reason, we also explored the effect of added electrolytes and solvent desiccation using the suspension from our previous work:[4]

(i) To understand how much the pitch can be reduced by the methods presented here and if it can lead to visible colouration, the ionic strength of the previously reported CNC suspension[4] was increased from 74 μmol $g^{-1}$ to 115 μmol $g^{-1}$ of sulfuric acid per CNC. This had the result of reducing the pitch within the microparticles from 1200-1400 nm (i.e. 1.2-1.4 μm) to 700-800 nm. However, further increasing the acid concentration to 212 μmol $g^{-1}$, was found to result in spheroidal particles (i.e. negligible buckling) that did not show significant self-organisation of the cholesteric phase during the drying process (data not shown). As such, while suspension optimisation can be used to significantly reduce the pitch, this alone is insufficient to reach visible wavelengths. Note for simplicity, the electrolyte added to the heat-treated suspension was sulfuric acid, but salt (such as NaCl) was seen to act similarly.

(ii) The effect of a solvent post-treatment was also explored. Microparticles prepared from the optimised formulation (i.e. 115 μmol $g^{-1}$) were washed with methanol and subsequently dried. This resulted in a reduced size with increased buckling across the surface. Moreover, very localised patches of red colour could be observed on some microparticles under the microscope. Cross-sectional SEM confirmed that the pitch had decreased to around 400-500 nm in the limbs, corresponding to reflection predominantly at infrared wavelengths (data not shown).

As such, while the approaches reported in this article can significantly reduce the pitch, it was not possible with the previous CNC suspension to produce visibly coloured microparticles.

Beyond the role of isolating the various directly accessible parameters, such as those exemplified above, the more fundamental parameters to consider are most likely the pitch before kinetic arrest and the volume fraction of CNCs at the onsets of both kinetic arrest and then buckling. These three quantities could, by themselves, be sufficient to describe the evolution of the pitch upon volume contraction until the formation of the final microparticles. While the pitch before kinetic arrest determines the overall starting position in the pitch *vs* volume fraction diagram (see Figure 5), the kinetic arrest transition defines where the pitch contraction suddenly decreases to $p \sim \Phi^{-1/3}$. As such, an early buckling transition, i.e., before most of the water is lost, is then essential to not only to restore a steeper pitch evolution, comparable to the $p \sim \Phi^{-1}$ we find in vertically aligned domains in dried films, but to allow sufficient drying time to exploit this pitch contraction and reach pitch values consistent with the visible range. Such parameters are likely related to the morphology of the CNC nanoparticles themselves and as such are likely related to the cellulose source and the extraction conditions used to yield the initial CNC suspension.

# Supplementary Tables

| Sample | $N$ (sample #) | $\mu$ / µm (mean) | $\sigma$ / µm (std dev) | $err_{\mu}$ / µm (SEM) | $\Delta\mu = \frac{(\mu_2 - \mu_1)}{\mu_1}$ | $err_{\Delta\mu}$ (SEM) |
|---|---|---|---|---|---|---|
| 1 - no methanol | 26 | 88.5 | 4.12 | ± 0.81 | -12.1% | ±1.1% |
| 2 - methanol | 20 | 77.9 | 2.26 | ± 0.50 | | |

**Supplementary Table 1.** Statistical analysis of the diameter Ø of the CNC microparticles from SEM analysis ($i$ = 1: no methanol), and its variation where a methanol treatment was previously applied ($i$ = 2: methanol). Here $\mu_i$ and $\sigma_i$ correspond to the mean and standard deviation of Ø in the sample $i$. The *standard error of the mean of* Ø (i.e., error on $\mu_i$) is estimated as $err_{\mu_i} = \sigma_i/N_i$, where $N_i$ is the number of measured particles. The variation of the mean diameter is calculated as $\Delta\mu = (\mu_2 - \mu_1)/\mu_1$, while $err_{\Delta\mu}$ quantifies its *standard error*, obtained by error propagation of their respective $err_{\mu_i}$ terms:

$$\frac{err_{\Delta\mu}}{\Delta\mu} = \sqrt{\frac{err_{\mu_2}^2 - err_{\mu_1}^2}{(\mu_2 - \mu_1)^2} + \frac{err_{\mu_1}^2}{\mu_1^2}}.$$

| Pigment colour | CNC / wt.% | [NaCl]/[CNC] / µmol g$^{-1}$ | Solvent post-treatment |
|---|---|---|---|
| red | 7.0 | 100 | none |
| green | 7.0 | 85 | isopropanol |
| blue | 7.0 | 100 | methanol |

**Supplementary Table 2.** Suspension formulation and microparticle treatment conditions to produce the red, green, and blue photonic CNC pigments reported in Figure 1 of the article. Note that while solvent treatment can result in large, discrete changes in the colour of the microparticles, the ionic strength of the initial formulation can also be used to fine-tune the final optical appearance, as was employed here to achieve a visibly green dispersion after isopropanol treatment (85 µmol g$^{-1}$), rather than cyan in appearance (100 µmol g$^{-1}$). The solvent post-treatment was not limited to those in this table, but validated for a wide range of polar solvents, including: methanol, ethanol, isopropanol, *t*-butanol acetone, acetonitrile etc.

# Supplementary Figures

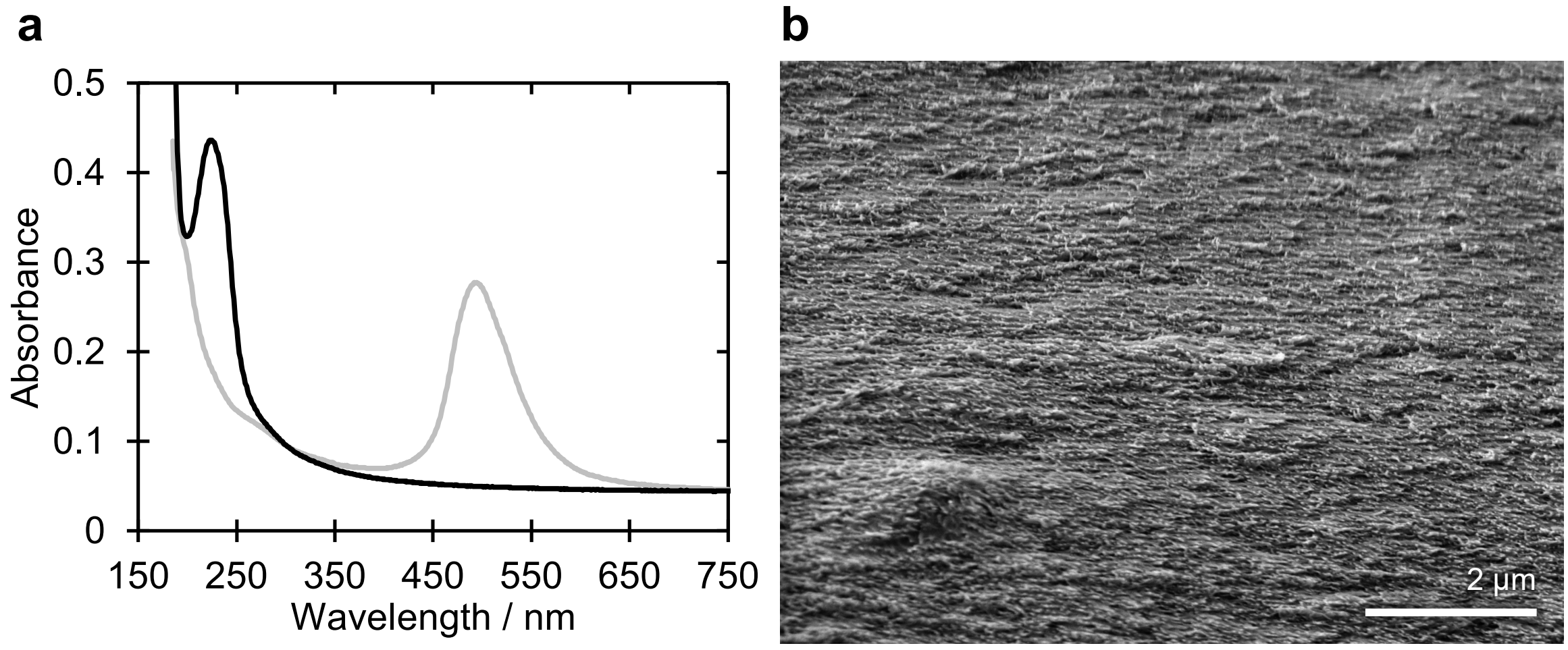

**Supplementary Figure 1.** Characterisation of a CNC film cast in a vertical magnetic field ($\mu_0 H \approx 0.6$ T) from a 50% dilution of the stock CNC suspension, i.e., the anisotropic phase of a commercial cellulose nanocrystal suspension (Univ. Maine, 7.0 wt.%) with [NaCl]/[CNC] ratio of 100 µmol $g^{-1}$. **(a)** The UV-vis spectrum of the resultant film (*black line*) shows a clear peak at $\lambda_{\mathrm{max}}$ = 224 nm that corresponds to a photonic response in the UVC region, and a strong absorption at <200 nm from cellulose itself, as validated against a second CNC film with visible colouration ($\lambda_{\mathrm{max}}$ = 493 nm, *grey line*). **(b)** The corresponding cross-sectional scanning electron microscopy (SEM) image of the UV film revealed a well-ordered helicoidal architecture with an mean pitch of 141 ± 9 nm (N = 30, uncertainty corresponds to standard deviation).

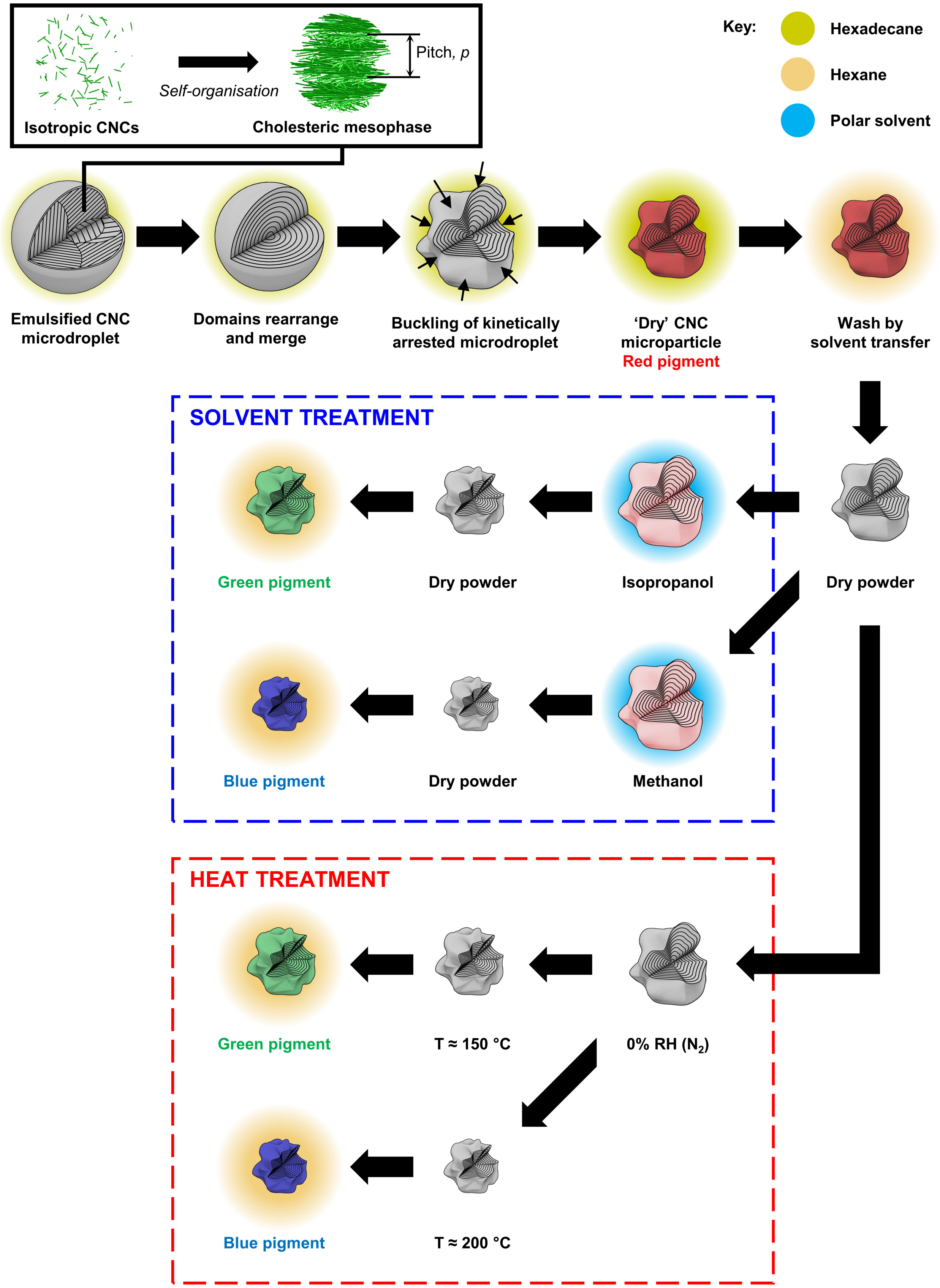

**Supplementary Figure 2.** An expanded schematic detailing the controlled thermal and solvent post-treatment steps to produce a full spectrum of photonic CNC pigments. The colour of the microparticle represents the typical colour reflected from the photonic nanostructure.

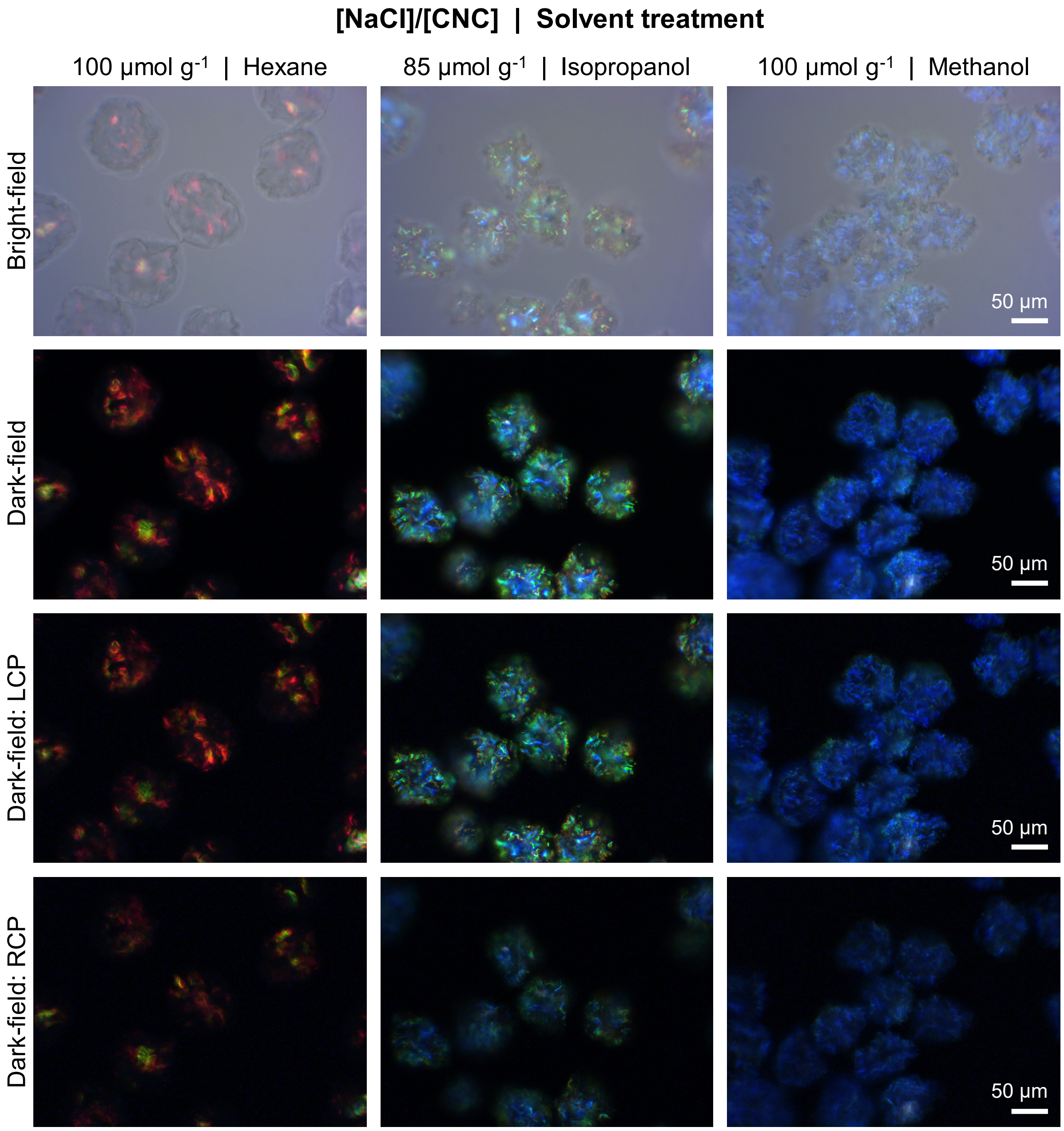

**Supplementary Figure 3.** Bright-field (BF) and dark-field (DF) optical microscope images of red, green, and blue microparticles (as specified in Supplementary Table 2) immersed in refractive index-matching oil ($n$ = 1.55). The reflected colour is less vivid in the bright-field images due to the buckling-induced misalignment of the CNC domains that do not reflect in the specular direction of illumination, combined with the dominant specular reflection from the glass substrate that is inherent to imaging from below using an inverted microscope. The microparticles were also imaged through left-handed circularly polarisation (LCP) and right-handed circularly polarisation (RCP) filters, showing predominant reflection in the left channel.

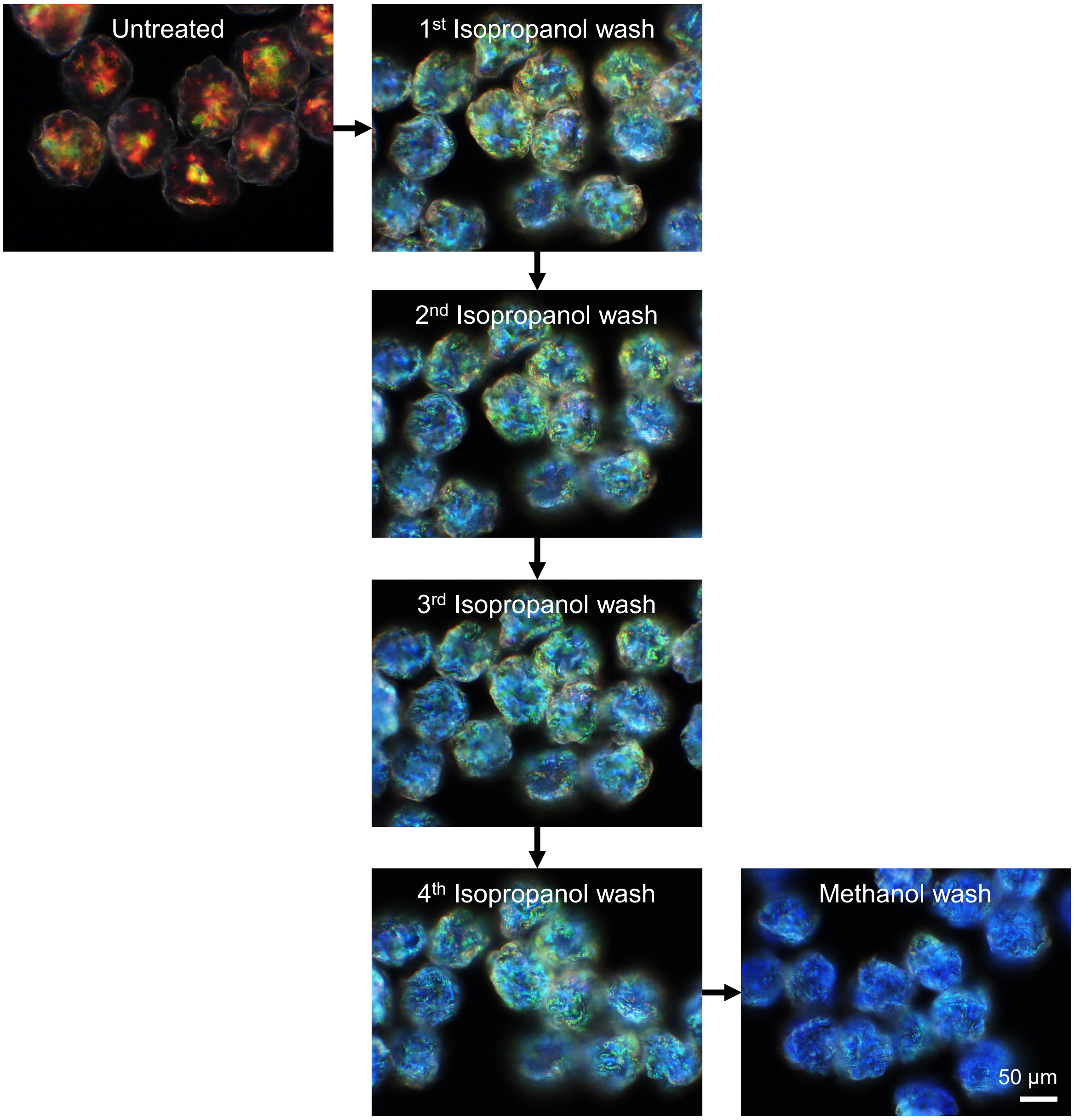

**Supplementary Figure 4.** Optical micrographs showing the change in the optical appearance of CNC microparticles upon successive treatments with isopropanol, and after a final treatment with methanol, a more polar solvent. After each solvent treatment, the sample was dried and immersed in non-polar hexane to remove surface scattering during imaging, as exemplified in Figure 2a-e of the article. After the first wash, additional treatments with isopropanol did not induce any significant changes to the reflected colour. However, treatment with methanol could still induce a further blueshift.

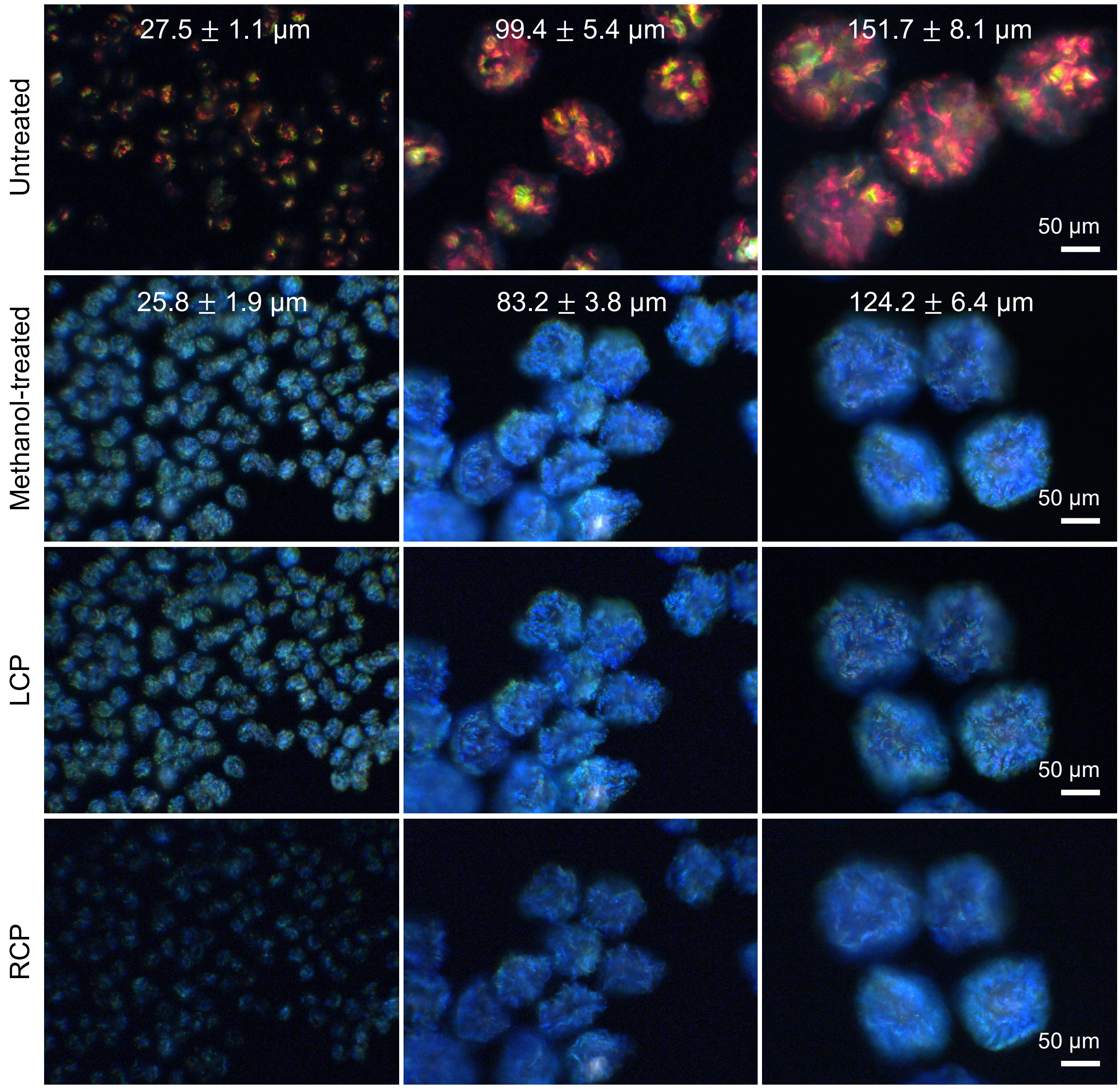

**Supplementary Figure 5.** Dark-field microscope images of three sizes of CNC microparticles recorded prior to, and after methanol treatment. Imaging the latter through left-handed circularly polarisation (LCP) and right-handed circularly polarisation (RCP) filters only permits light with the same polarisation to be captured, allowing the ordering of the internal architecture to be assessed.

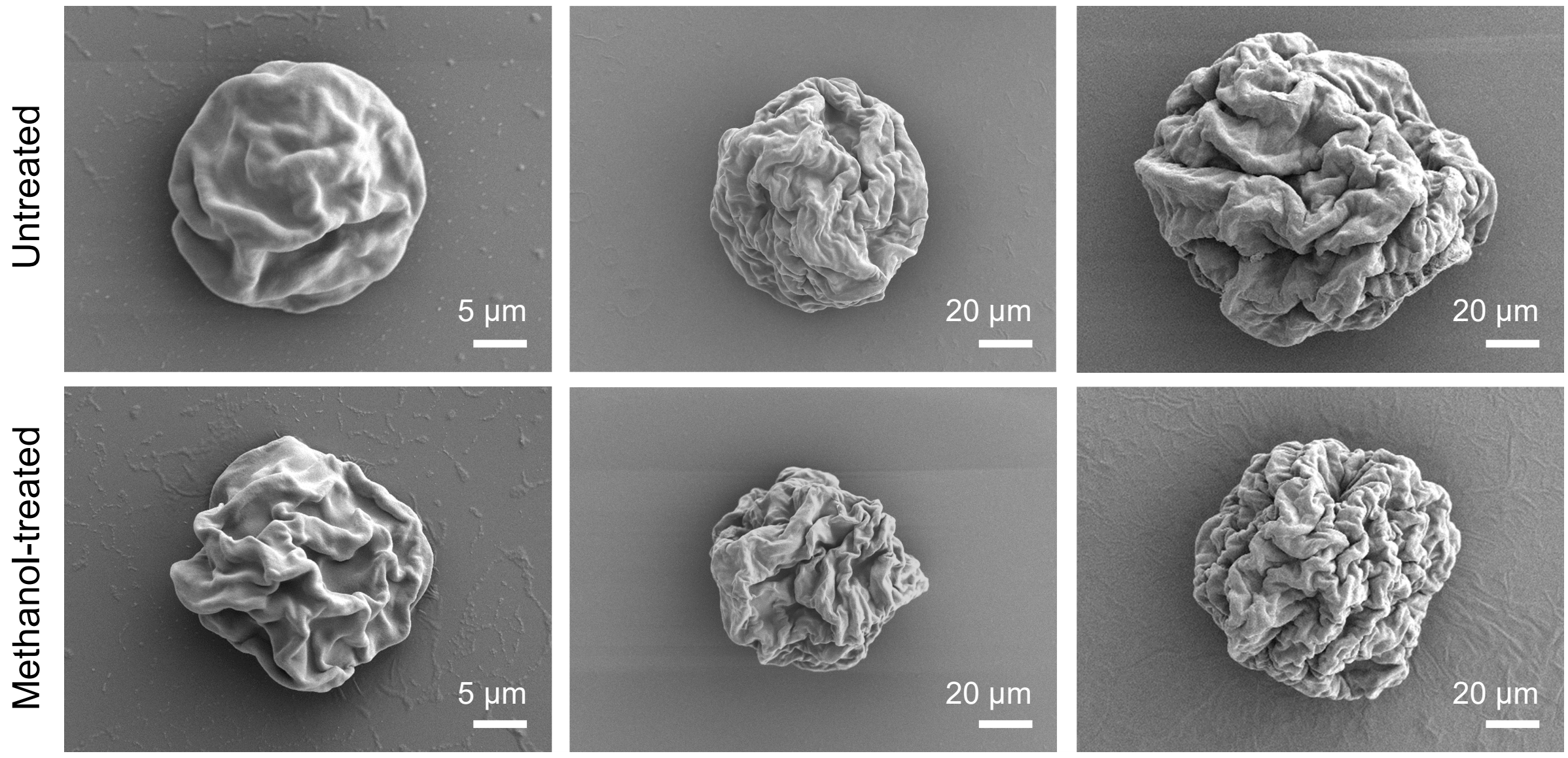

**Supplementary Figure 6.** Scanning electron microscope images of the exterior of a representative CNC microparticle of each of the three size regimes reported in Supplementary Figure 5, prior to (*top*) and after post-treatment with methanol (*bottom*). A decrease in diameter with a corresponding increase in surface buckling is apparent in all three cases.

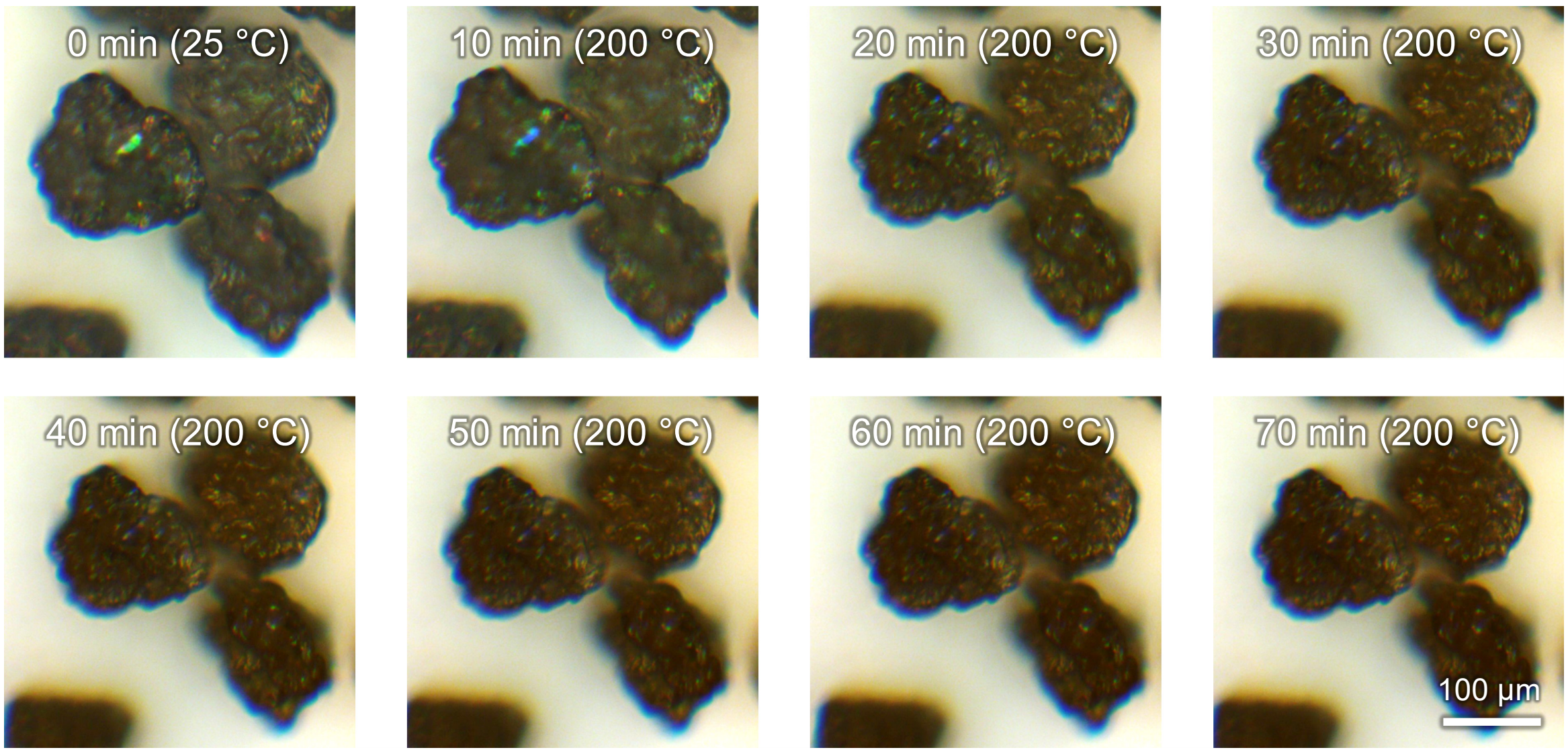

**Supplementary Figure 7.** Time-lapse reflectance microscope image series showing the decrease in the size of the CNC microparticle during heat treatment at 200 °C for one hour, with most of the shrinkage occurring during the first thirty minutes. The heating ramp rate was set to 20 °C min$^{-1}$.

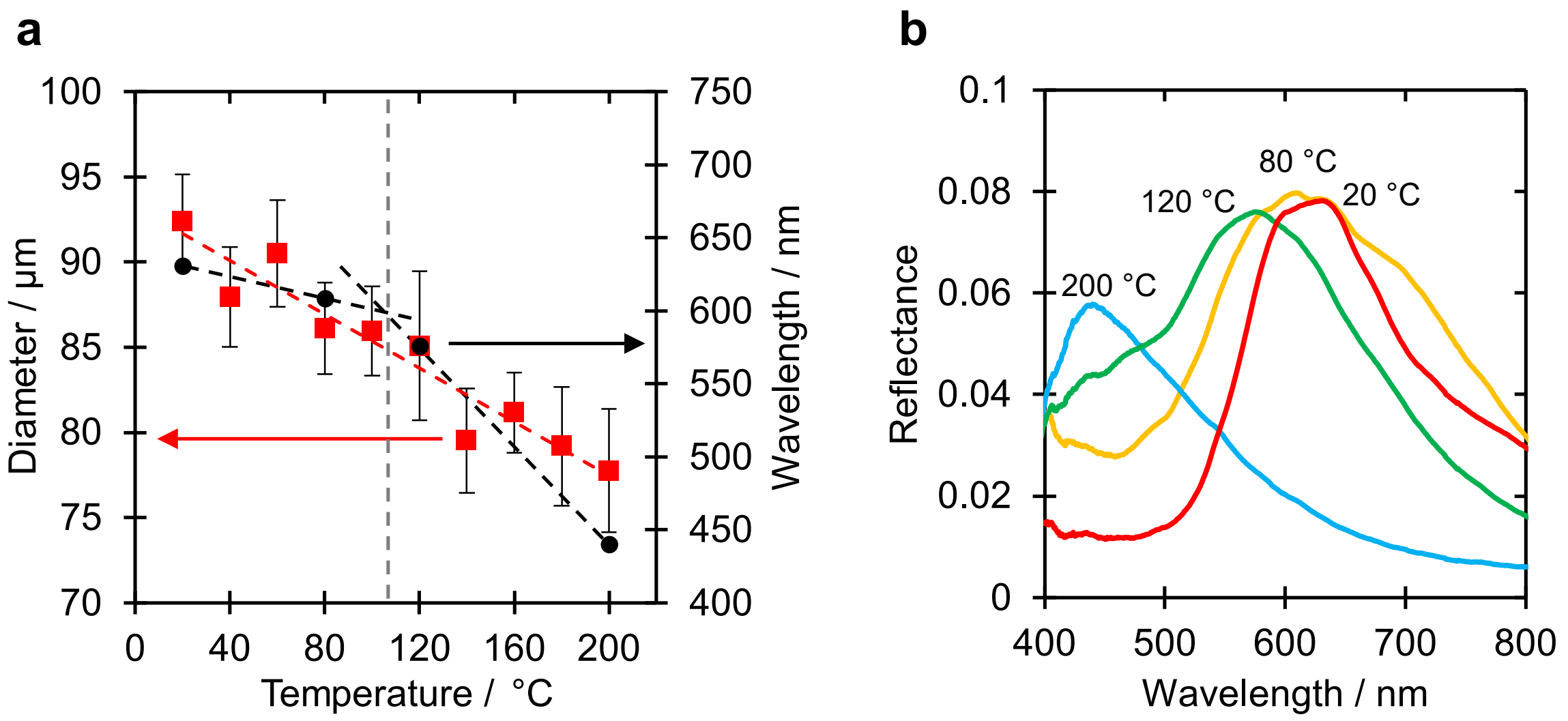

**Supplementary Figure 8. (a)** The linear reduction in CNC microparticle diameter upon thermal treatment (*red squares*) and the corresponding peak wavelength (*black circles*), showing the significant blueshift above 100 °C. **(b)** Representative spectra for the heat-treated CNC microparticles highlighted in (a), showing the increasing blueshift with temperature. Note that the decrease in reflected intensity at 200 °C, compared to e.g., methanol treatment (Figure 1d of the article), is attributed to the onset of thermal degradation. This is evidenced by the discolouration of the microparticles, which is apparent when imaged in the dry state (Supplementary Figure 9).

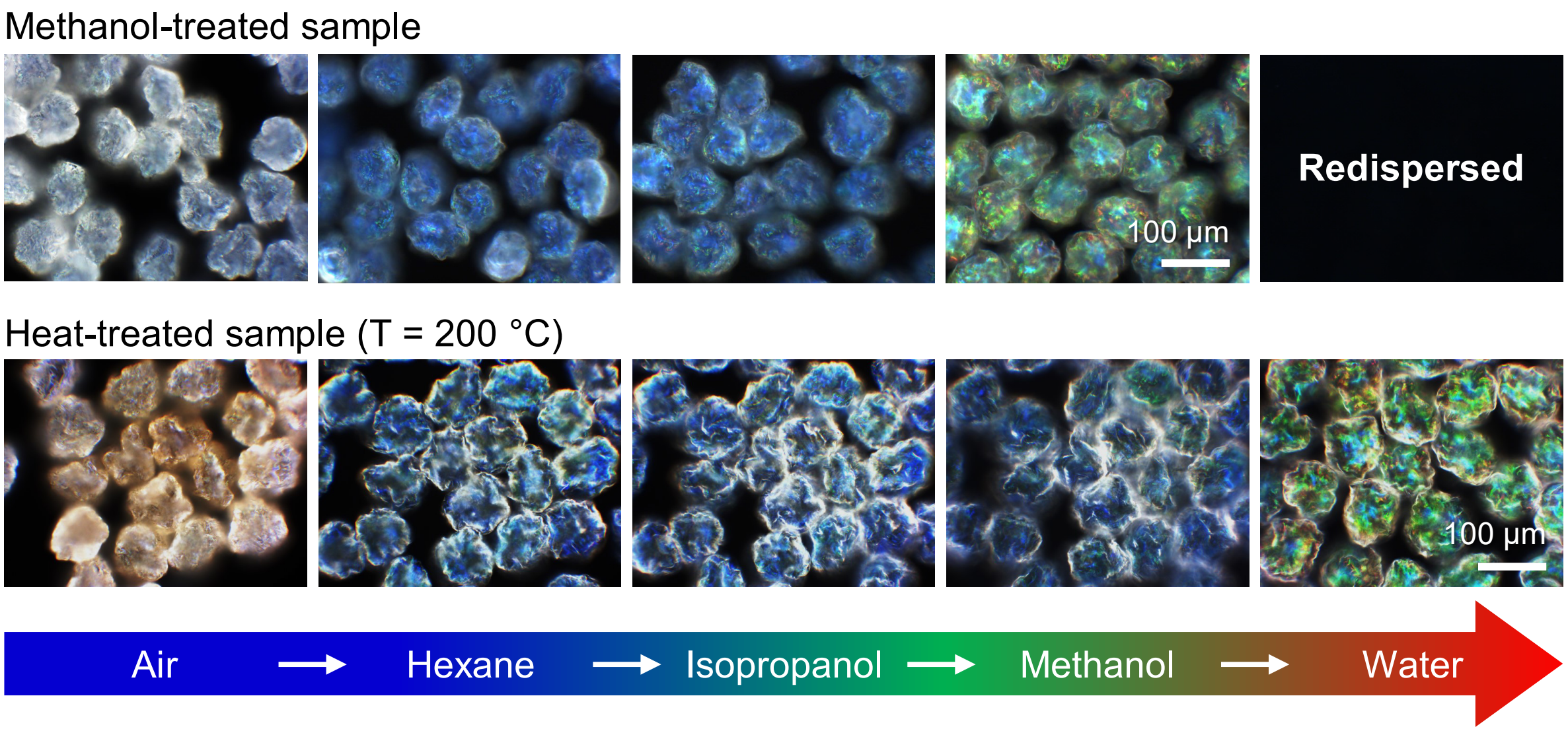

**Supplementary Figure 9.** Dark-field microscope images showing the responsiveness of the CNC microparticles to different solvents after post-treatment with either methanol (*top*) or by heating at 200 °C for 1h (*bottom*). Polar solvents swell and redshift the methanol-treated microparticles, with water able to completely disperse the methanol-treated sample. In contrast, the 200°C heat-treated sample did not swell as extensively and was stable in water.

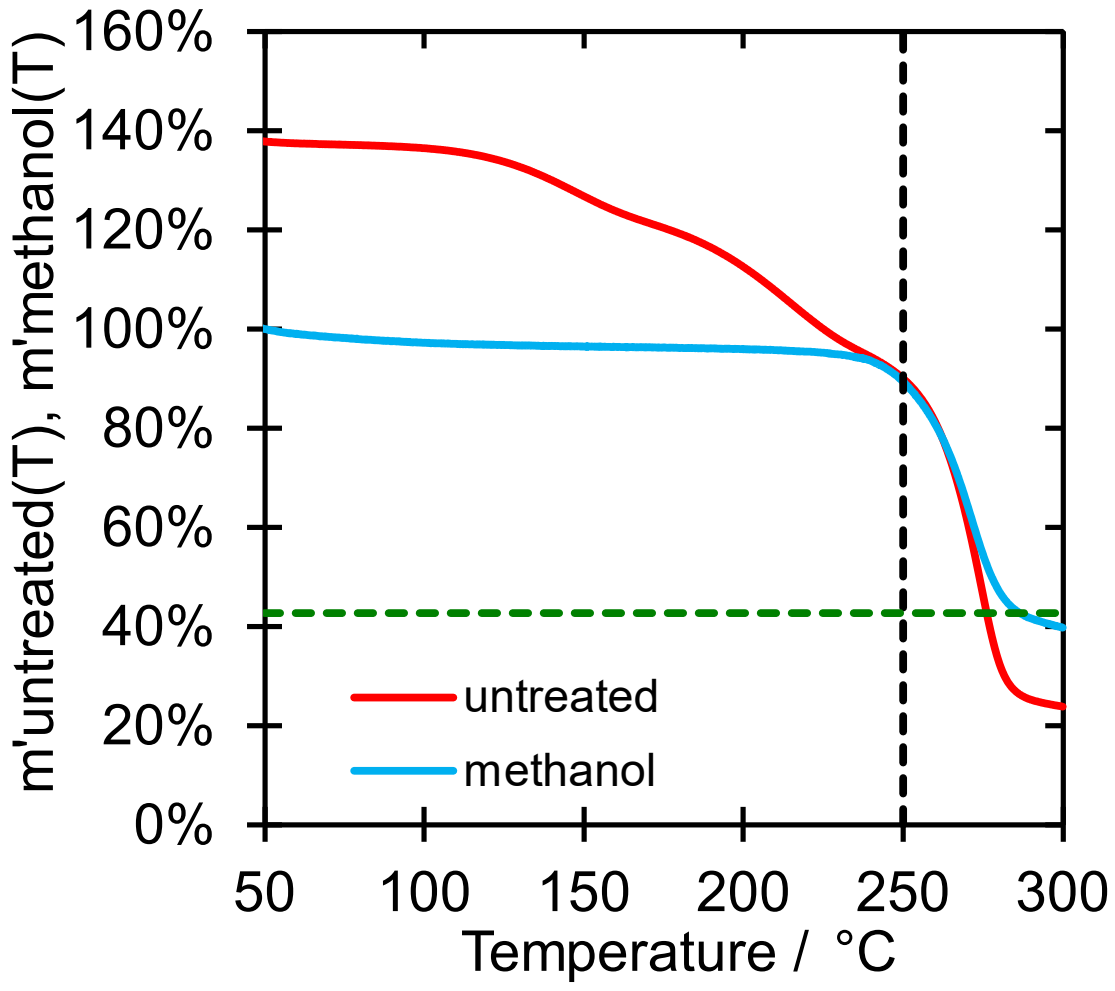

**Supplementary Figure 10.** Processed thermogravimetric analysis (TGA) curves of the microparticles comparing untreated microparticles (i.e., hexane-washed only, *red line*) with methanol-treated microparticles (*blue line*). The mass of the untreated sample was rescaled as $m'_{\text{untreated}}(T) = \text{TGA}_{\text{untreated}}(T)\, w_{\text{CNC},2}/w_{\text{CNC},1}$, to account for the same initial cellulose content such that it overlaps at 250 °C with the methanol-treated sample (*vertical dashed line*). This rescaling makes the residual water in the untreated microparticles more apparent, but also indicates that at 300 °C, methanol treatment significantly prevents mass loss *via* pyrolysis, approaching the theoretical maximum (*horizontal dashed line* at $42.67\% = r_{(\text{C/CNC}),\max} w_{\text{CNC},2}$, see Supporting Discussion 1 for details). This agrees with the dehydrant property of methanol,[1] as the removal of water (and thus oxygen atoms) prevents mass loss *via* CO and $CO_2$ emission.

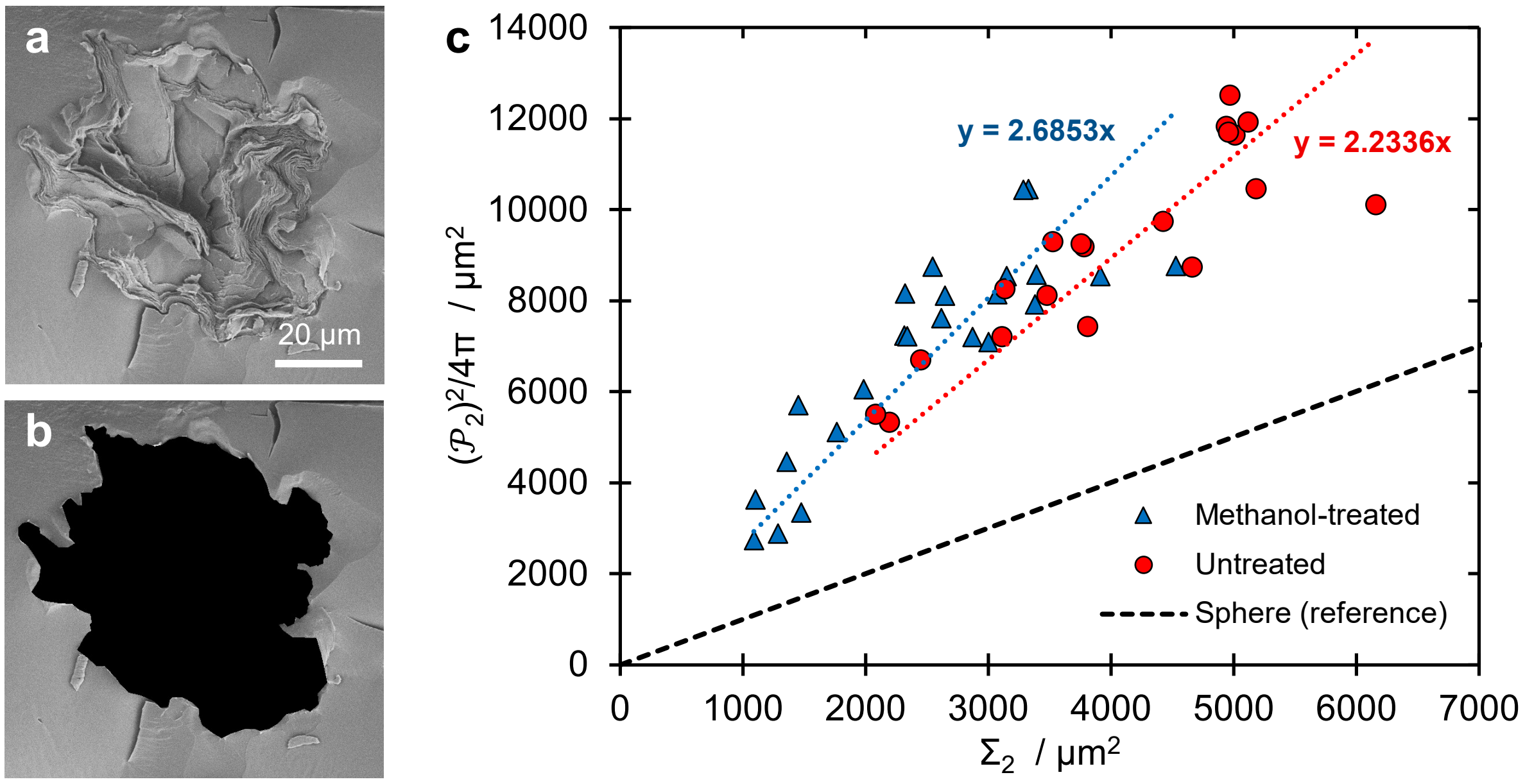

**Supplementary Figure 11. (a)** Example of cross-sectional SEM analysis of a methanol-treated CNC microparticle, illustrating how the buckled surface distorts the radially aligned internal helicoidal architecture. **(b)** The manually shaded area corresponding to the cross-section of the microparticle shown in (a), allowing for determination of both the apparent perimeter ($\mathcal{P}_2$) and apparent cross-sectional area ($\Sigma_2$) of the microparticle. **(c)** A plot showing the correlation between $\mathcal{P}_2$ and $\Sigma_2$ for both methanol-treated (*blue triangles*) and untreated (*red circles*) CNC microparticles with different fracture planes. The corresponding relationship for an unbuckled sphere is included for reference (*black dashed line*). The slope of the trendlines provides an estimation of $\beta$ and $Q$, given $Q = \beta^2 = 4\pi\Sigma_2/\mathcal{P}_2^2$.

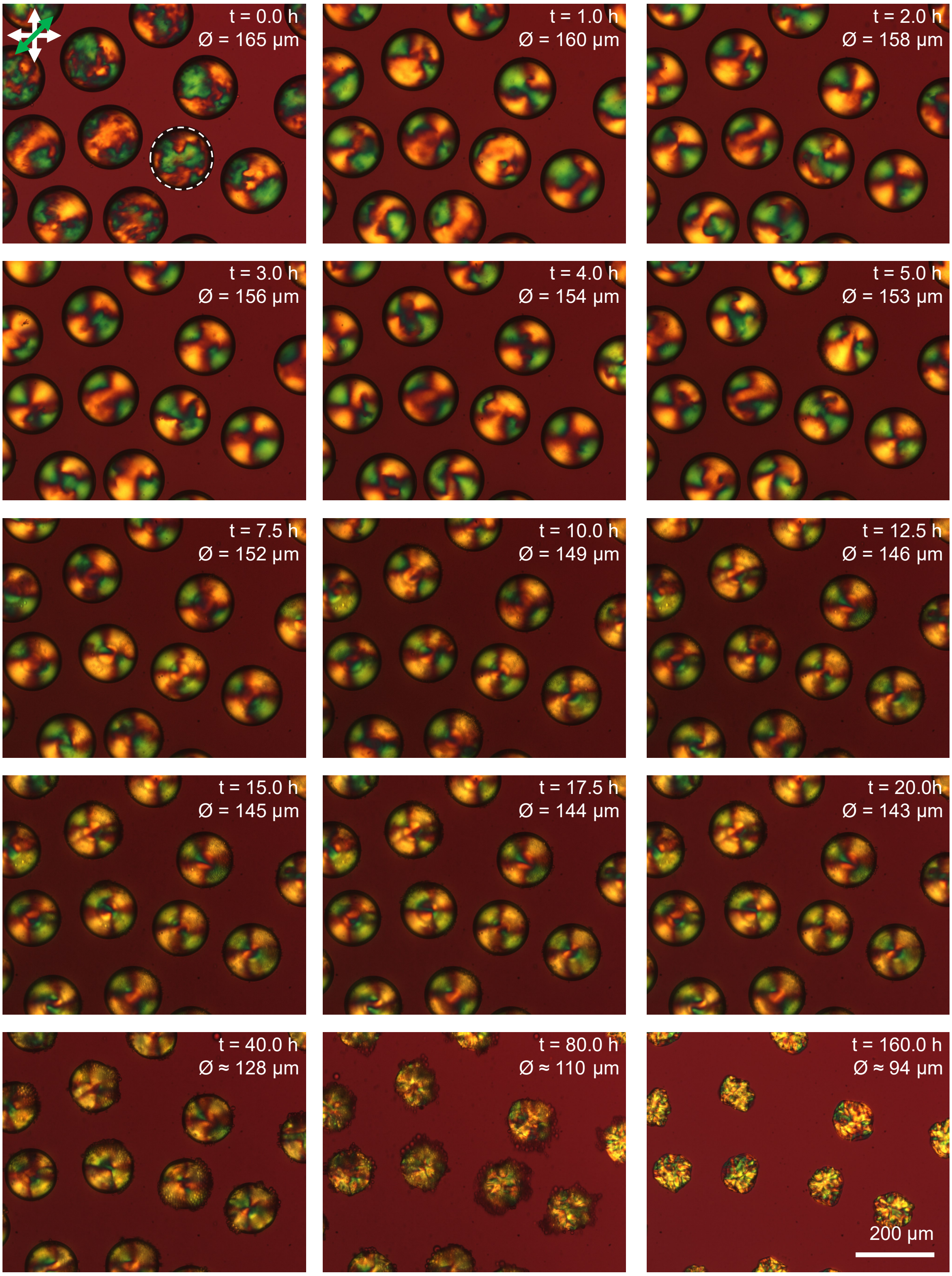

**Supplementary Figure 12.** Time-lapse polarised optical microscope image series showing the evolution of a drying CNC microdroplet ([CNC] = 7 wt.%, [NaCl]/[CNC] = 100 μmol $g^{-1}$, anisotropic phase). The micrographs were collected in transmission through crossed polarisers with a 45° full-wave retardation plate, allowing visualisation of the birefringent cholesteric phase. The yellow/blue colours indicate the mesophase orientation. The reduction in the drying rate and the cloudy appearance is due to localised saturation of the oil, resulting in surfactant-stabilised water droplets.

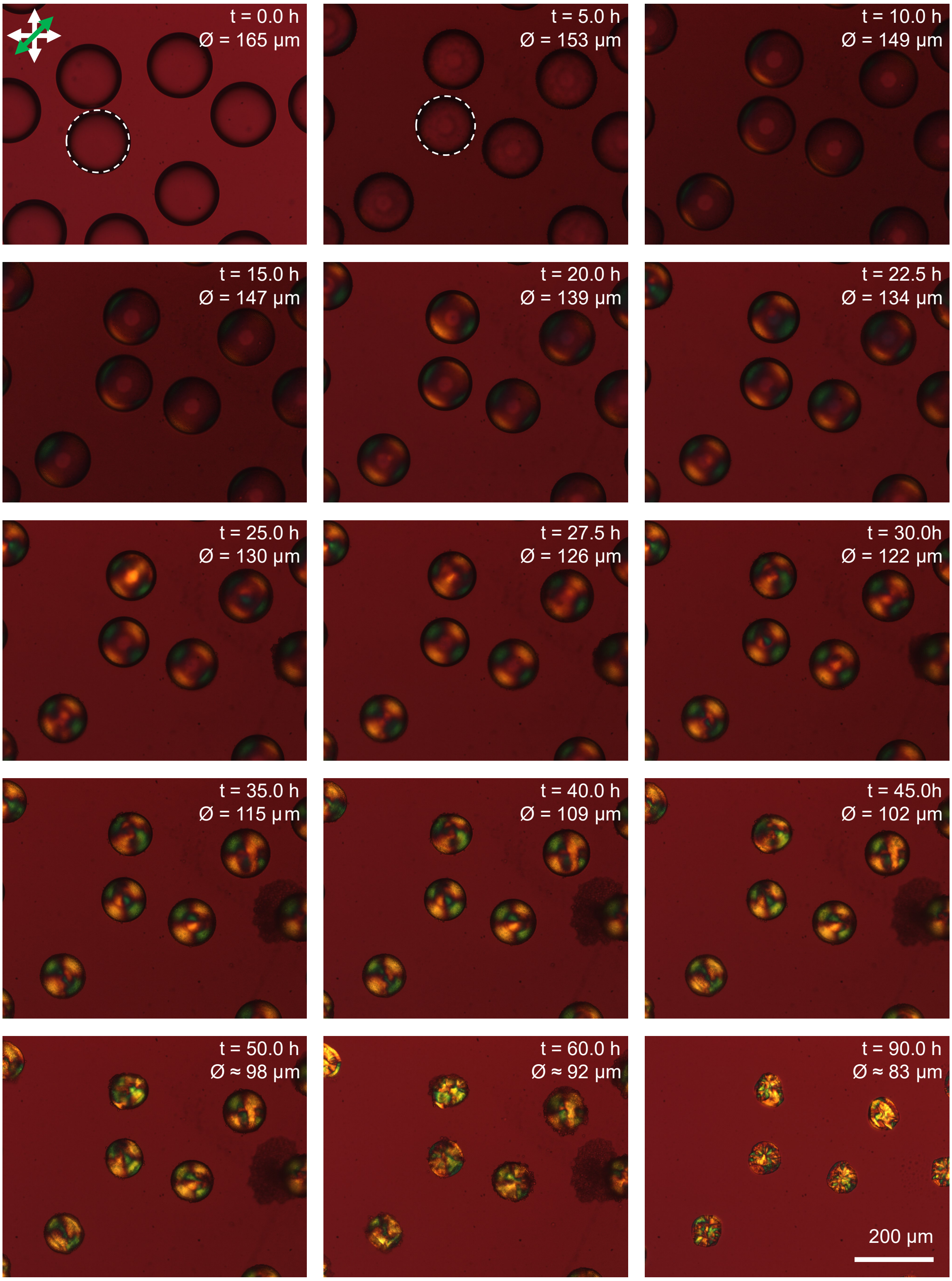

**Supplementary Figure 13.** Time-lapse polarised optical microscope image series showing the evolution of a drying CNC microdroplet ([CNC] = 3.5 wt.%, [NaCl]/[CNC] = 100 µmol g$^{-1}$, i.e., 50% dilution of the stock anisotropic suspension). The micrographs were collected in transmission through crossed polarisers with a 45° full-wave retardation plate, allowing visualisation of the birefringent cholesteric phase. The yellow/blue colours indicate the mesophase orientation. The reduction in the drying rate and the cloudy appearance is due to localised saturation of the oil, resulting in surfactant-stabilised water droplets.

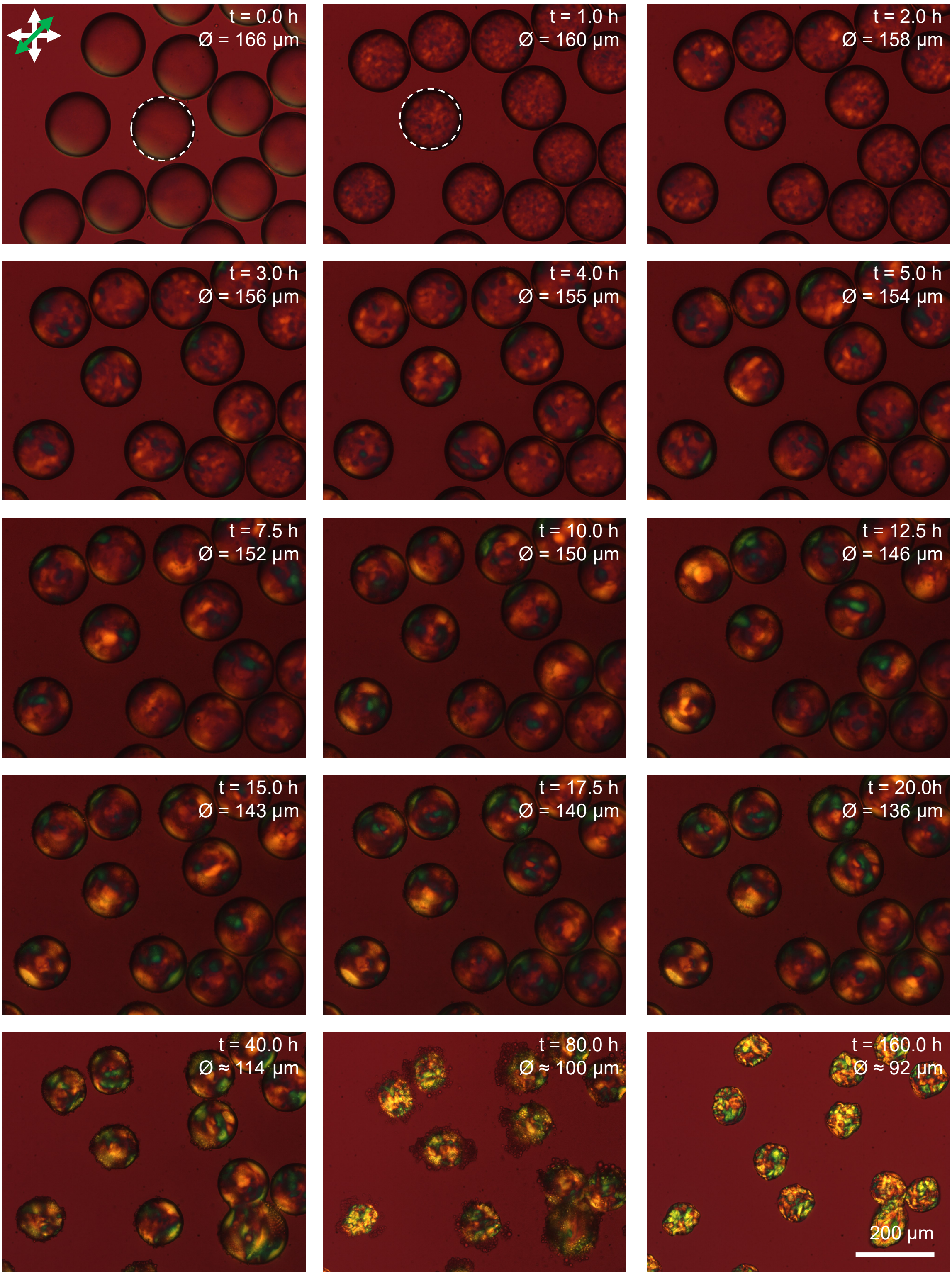

**Supplementary Figure 14.** Time-lapse polarised optical microscope image series showing the evolution of a drying CNC microdroplet ([CNC] = 7 wt.%, [NaCl]/[CNC] = 100 μmol $g^{-1}$, isotropic phase). The micrographs were collected in transmission through crossed polarisers with a 45° full-wave retardation plate, allowing visualisation of the birefringent cholesteric phase. The yellow/blue colours indicate the mesophase orientation. The reduction in the drying rate and the cloudy appearance is due to localised saturation of the oil, resulting in surfactant-stabilised water droplets.

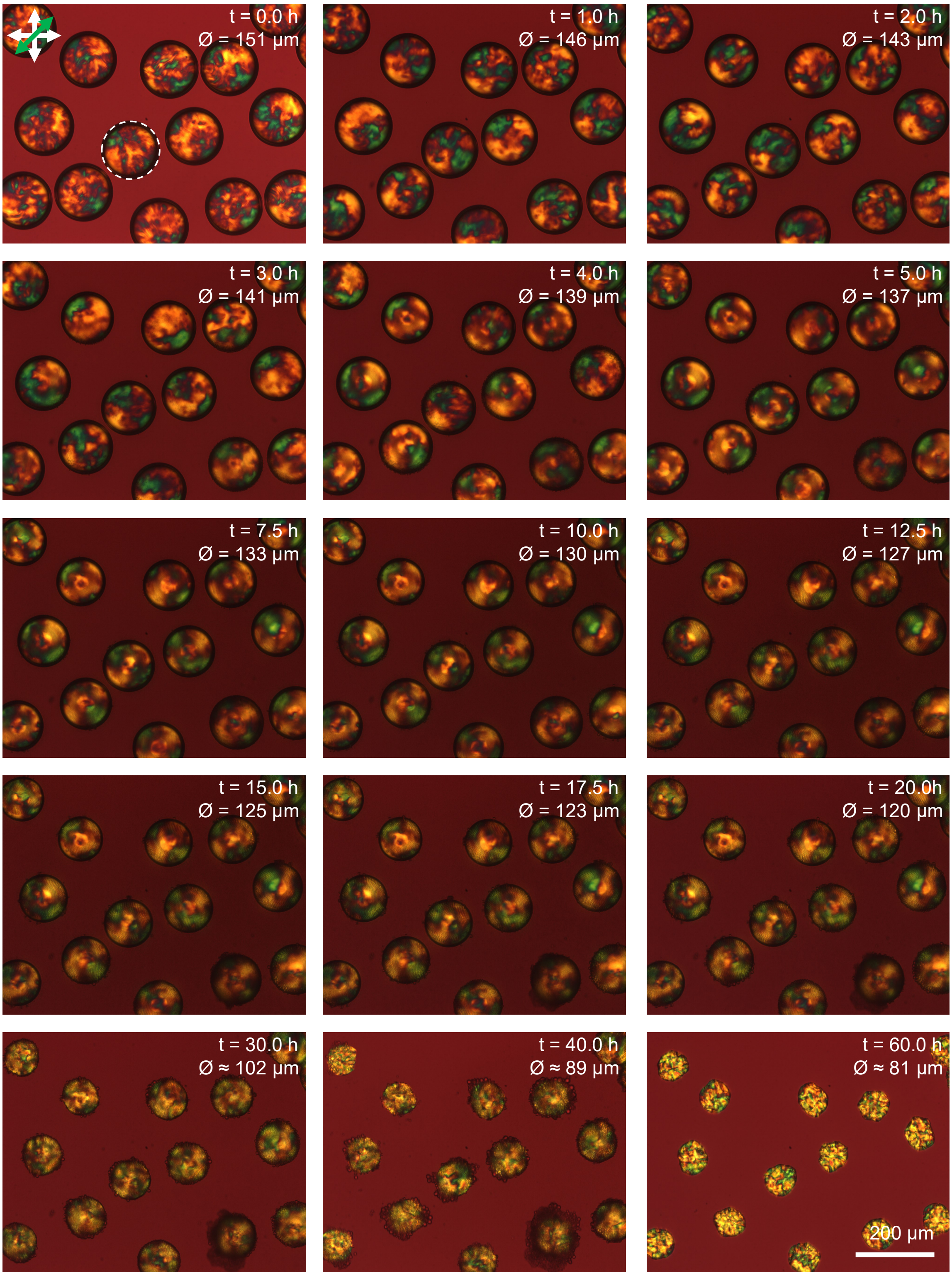

**Supplementary Figure 15.** Time-lapse polarised optical microscope image series showing the evolution of a drying CNC microdroplet ([CNC] = 7.0 wt.%, [NaCl]/[CNC] = 0 µmol $g^{-1}$, i.e., no added salt). The micrographs were collected in transmission through crossed polarisers with a 45° full-wave retardation plate, allowing visualisation of the birefringent cholesteric phase. The yellow/blue colours indicate the mesophase orientation. The reduction in the drying rate and the cloudy appearance is due to localised saturation of the oil, resulting in surfactant-stabilised water droplets.

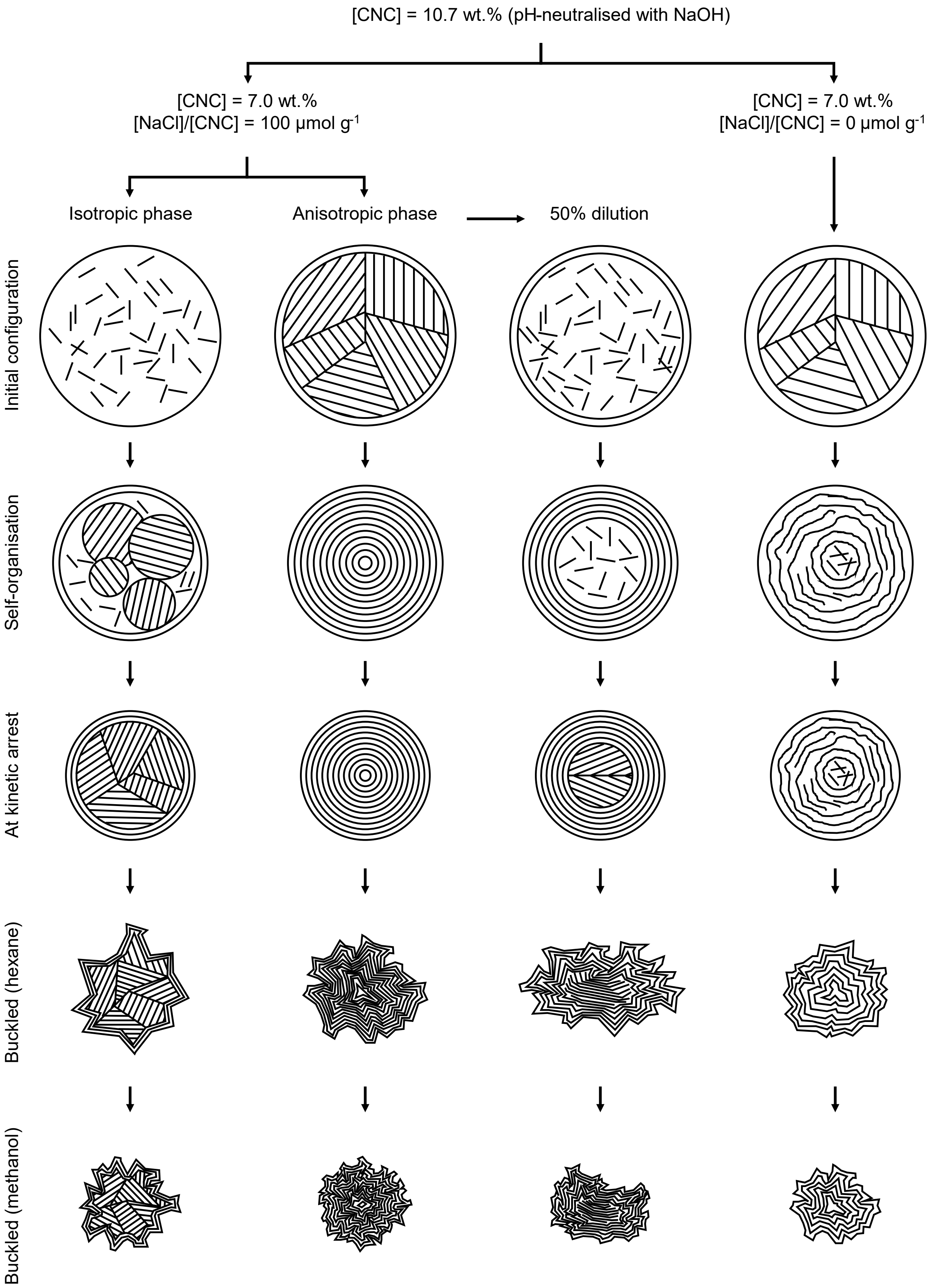

**Supplementary Figure 16**. Schematic describing the self-assembly pathways for the different formulations of the CNC suspension, including the effect of solvent desiccation as a post-treatment. The morphology is derived from the time-lapse studies reported in Supplementary Figures 12-15 and the optical and electron microscopy in Figure 6 of the article.

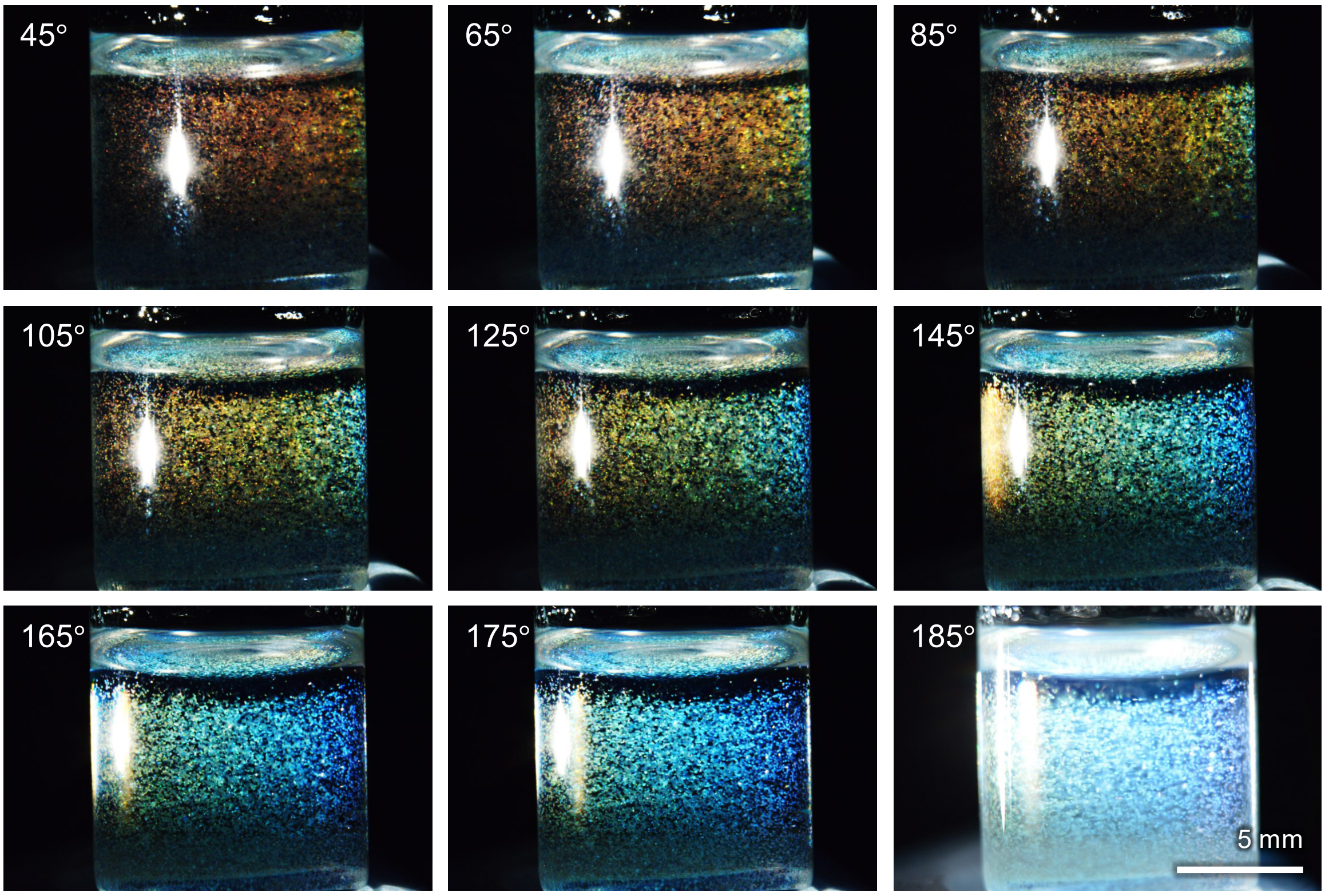

**Supplementary Figure 17.** Photograph sequence showing the iridescence of red CNC microparticles dispersed in refractive index matching oil ($n = 1.55 \approx n_{\mathrm{CNC}}$) within a cylindrical glass vial, when viewed at 0° and with the illumination angle of a directed white light beam (3300 K) increased from $\alpha_i$ = 45° to 180° (as defined with respect to the viewing direction). This geometry allows the incident and outgoing light beams inside the suspension to remain relatively unaffected by Snell's law at the air/glass and glass/oil interfaces. In this configuration, the particles showed a drastic colour change from red to blue with increasing $\alpha_i$.

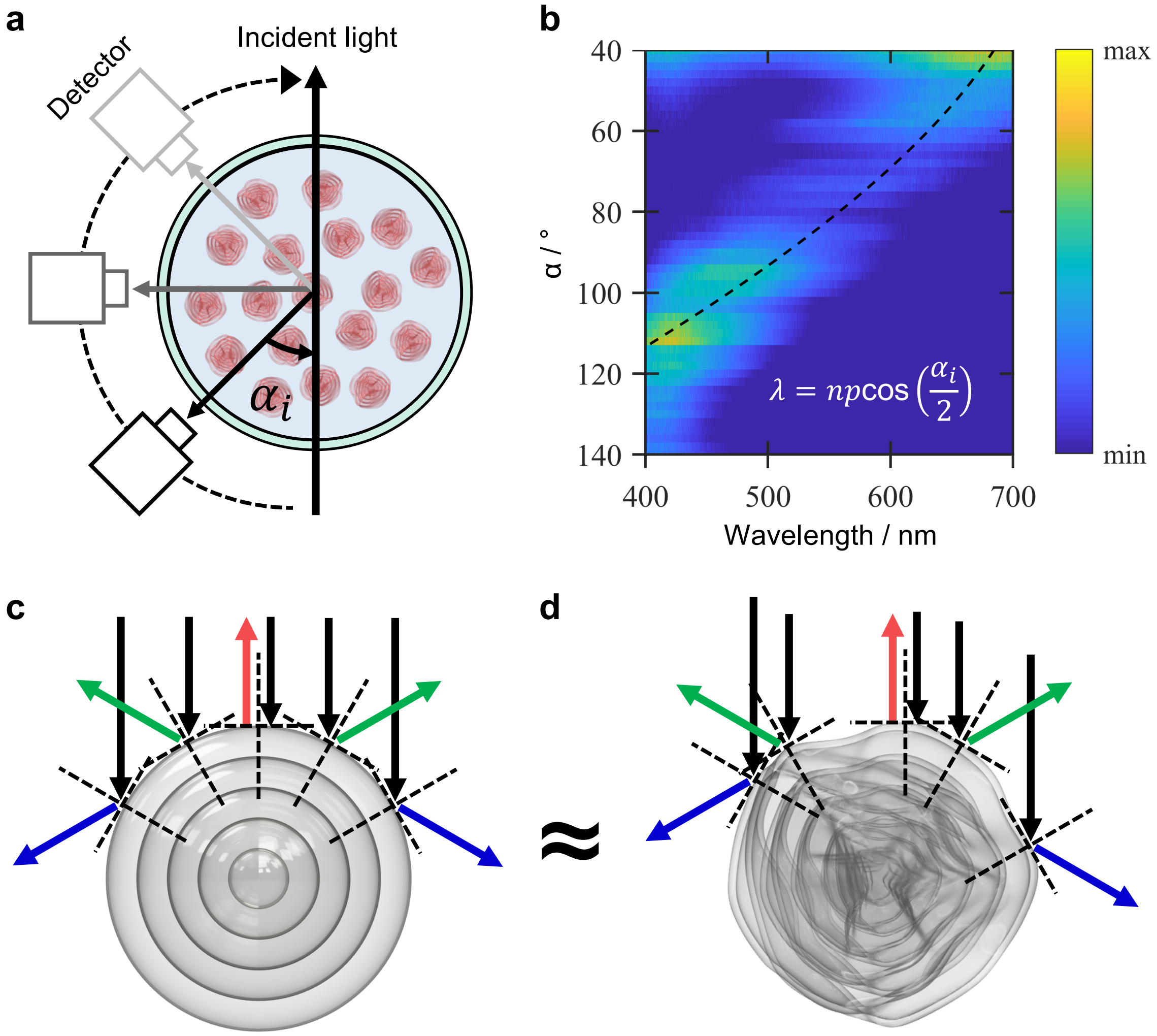

**Supplementary Figure 18. (a,b)** Angular resolved spectrometry of the red CNC microparticles dispersed in refractive index matching oil ($n = 1.55 \approx n_{\mathrm{CNC}}$) within a cylindrical glass vial (see Supplementary Figure 17), where the sample and illumination direction are fixed, and the detector is swept over half a turn (equivalent to $\alpha_i$ = 45° to 180°). The peak reflection shifts according to Bragg's law and as such the microparticles can be considered to be always at specular conditions. This optical behaviour matches surprisingly well with the expected isotropic response for a radially aligned cholesteric sphere,[5,6] despite the presence of buckling. This confirms that buckling has minimal impact on the macroscopic angular optical response of the CNC microparticles. **(c,d)** A simplistic model explaining the similarity between the optical response of (c) the non-buckled microparticles and (d) buckled microparticles under these illumination conditions.

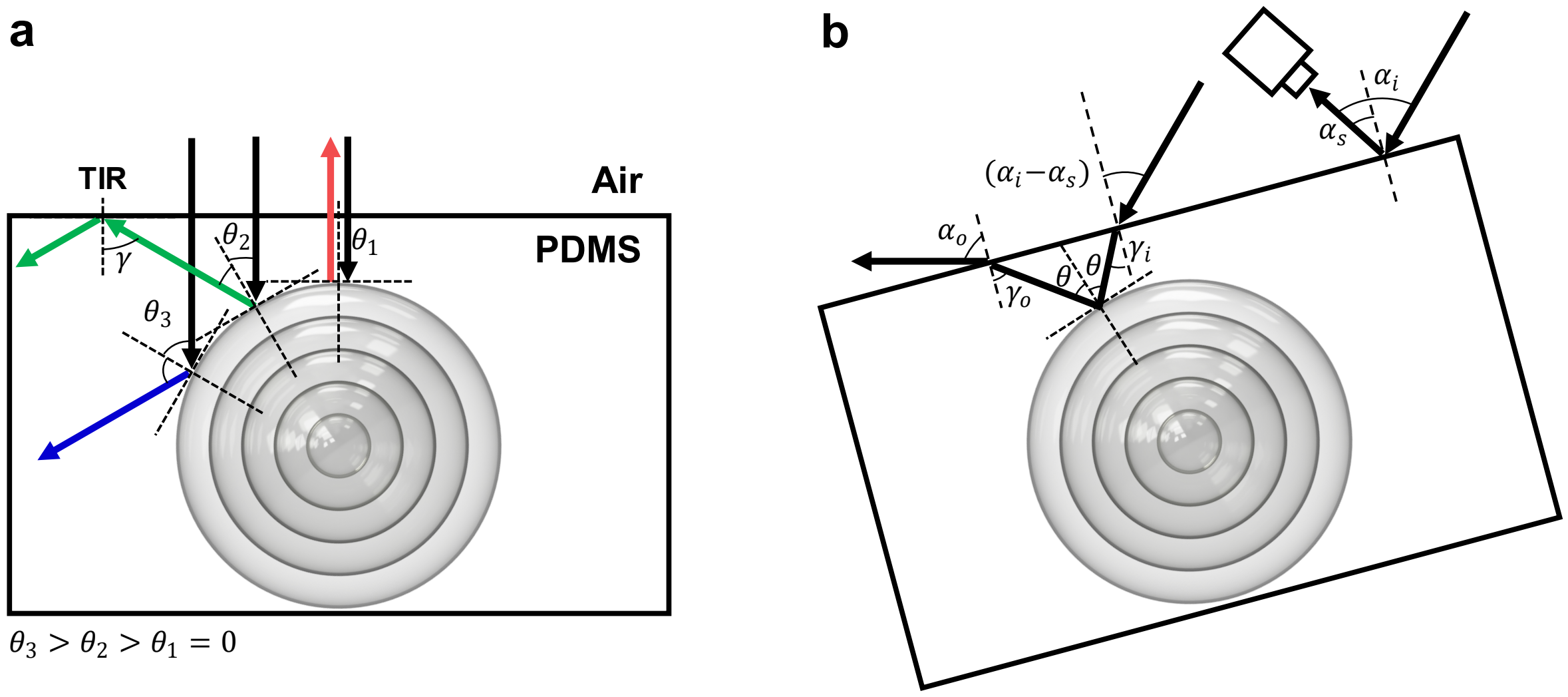

**Supplementary Figure 19. (a)** Schematic illustrating the conditions required for light reflected from an embedded CNC microparticle to escape from the PDMS matrix, for the simple case where the incident light is perpendicular to the air/film interface. **(b)** A more complicated situation where the sample angles and illumination angles are at arbitrary values.

When considering broadband incident light perpendicular to the film surface, as in (a), refraction at the resin-air interface does not occur. Among the beams that impinge a given microsphere, those with incident angle $\theta > 45°$ are reflected towards the back of the film (e.g., blue arrow). Light with incident angle $\theta < 45°$ arrives at the resin-air interface at an angle $\gamma = 2\theta$. If this angle $\gamma$ is larger than the critical angle ($\sim 44.3°$ for PDMS resin with $n_{\text{PDMS}}$ = 1.43), total internal reflection occurs according to Snell's law and the light cannot escape the resin (e.g., green arrow). Therefore, the maximum $\theta$ = $44.3°/\,2$ = 22.1°. The light incident on to the CNC microsphere will refract further to reduce the incident angle ($20.3°$), making the ratio between the detectable minimum and maximum wavelength: $\cos(20.3°) \approx 0.94$, which approximates to a maximum wavelength shift of 30 nm for the green CNC photonic pigment. In the more complex case, as illustrated in (b), the relationship between different angles are as following:

$$\gamma_i = \arcsin\left(\frac{1}{n_{\text{PDMS}}}\sin\left(\alpha_i - \alpha_s\right)\right)$$

$\theta$: variables - anything between (0, 90), determining the colour by $\lambda = np\,\cos\left(\text{arcsin}\left(\frac{n_{\text{PDMS}}}{n}\sin\theta\right)\right)$

$$\gamma_o = 2\theta - r_i$$

$$\alpha_o = \arcsin(n\sin(\gamma_o))$$

where $n = n_{\text{CNC}} = 1.55$, $\alpha_i$ is the angle between the incident light and the film surface, $\alpha_s$ is the angle between detection and film surface, and $\alpha_o$ is the angle between the refracted light beam and the film surface.

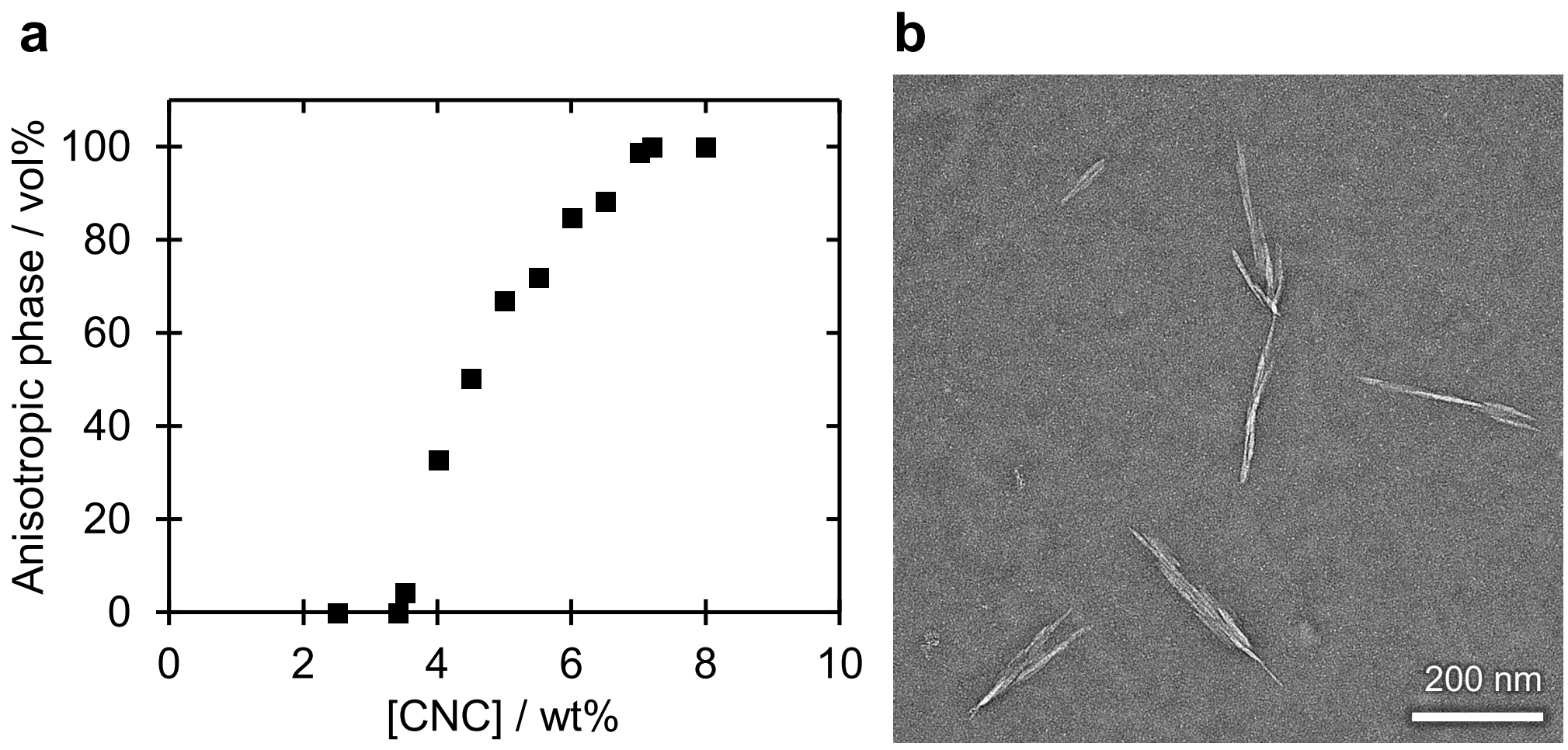

**Supplementary Figure 20. (a)** Volumetric ratio of anisotropic phase as a function of the concentration of the commercial CNC suspension (i.e., no added salt), as extracted from photographs of glass capillaries between crossed polarisers. **(b)** Transmission electron micrograph (TEM) of individual CNCs, showing their elongated, splinter-like morphology.

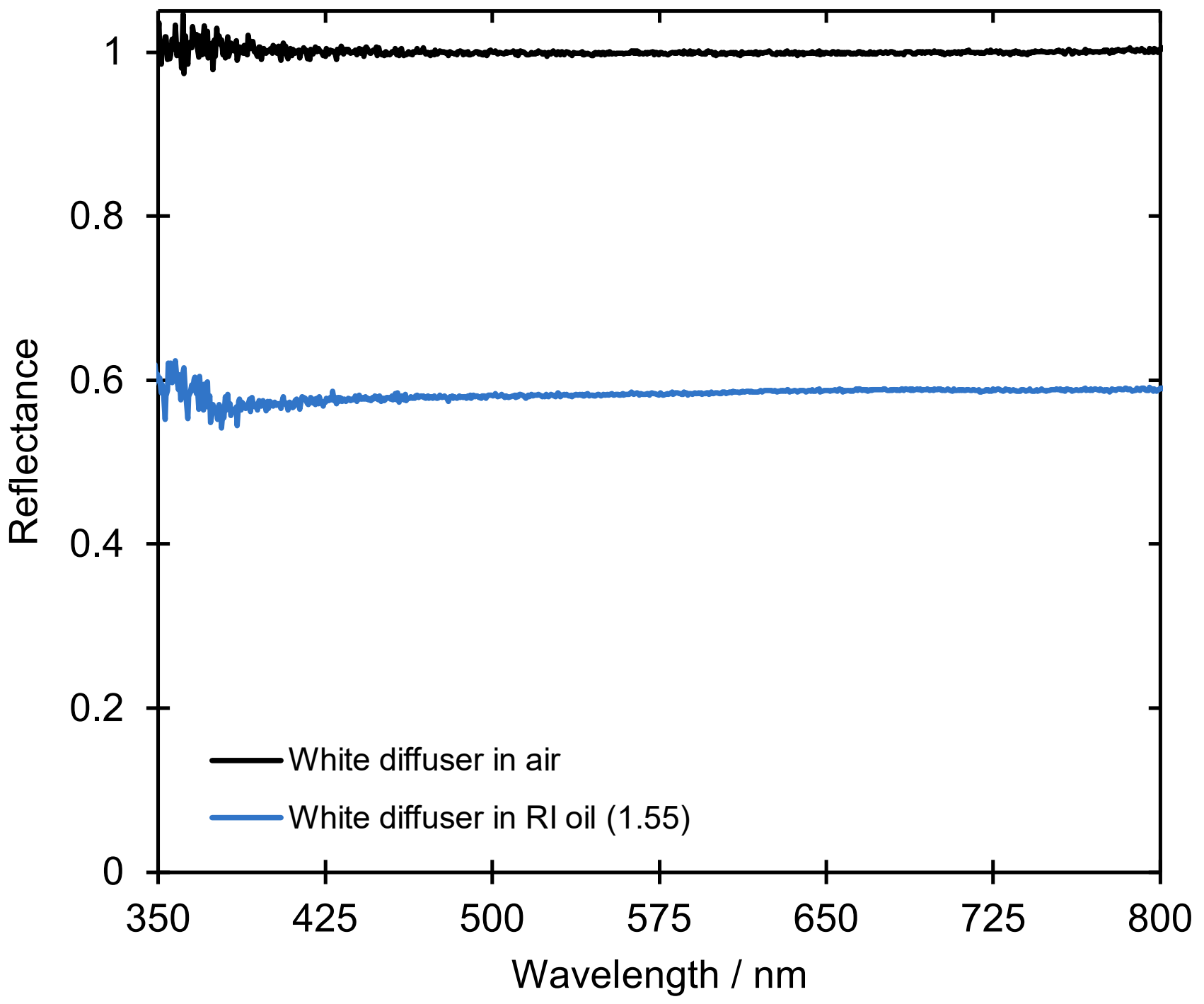

**Supplementary Figure 21.** Comparison of the reflectance from a white diffuser in dark-field illumination either (i) in air (*black line*) or (ii) coated with a thin droplet of refractive index oil ($n$ = 1.55, *blue line*), both spectra are normalised against the same white diffuser in air. The microscope configuration is the same as that used to analyse the CNC microparticles. By referencing the optical response of the CNC microparticles against a white diffuser coated with oil the specular reflection from the air-oil interface can be negated. However, it should be noted that as the optical response of the CNC microparticles is not based upon scattering, but arises from a distorted, radially aligned cholesteric ordering, the spectra reported in Figure 1 of the article may still underrepresent the true reflectivity of the pigments.[7]

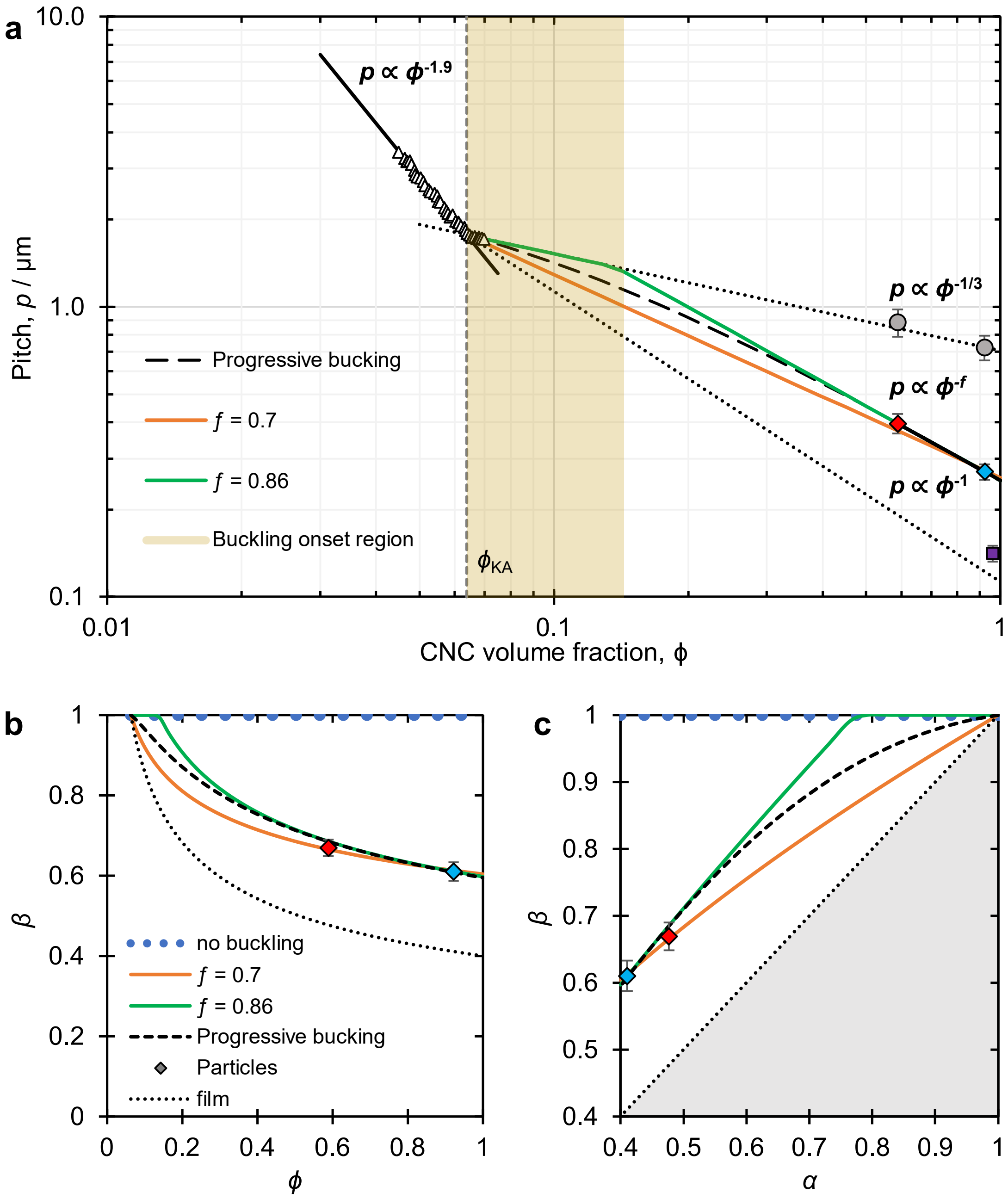

**Supplementary Figure 22. (a)** Pitch diagram describing the different compression scenarios after the onset of kinetic arrest: (i) vertical domain in a planar geometry, (ii) no buckling in a spherical geometry, (iii) power law $\phi^{-f}$ with $f$ = 0.7 when assuming a power law between $\beta$ and $\alpha$, (iv) delayed buckling with an initial unbuckling behaviour, followed by a power law $f$ = 0.86, assuming a power law between $\beta$ and $\alpha$, (v), a progressive buckling occurring at the onset of kinetic arrest and developing at higher volume fractions. The onset of buckling must be between the extremes of iii and iv, as denoted by the brown region. Coloured diamonds represent the particle pitch (from SEM) vs volume fraction (from TGA). Error bars represent the average deviation of $p_{\text{limb}}$ from SEM cross sections **(b)** Scaling laws between $\beta$ and $\phi$ for the different scenarios, and **(c)** the corresponding scaling law between $\beta$ and $\alpha$ (with the shaded triangle corresponding to unphysical $\beta$ values outside the [$\alpha$, 1] range). In (b) and (c) the coloured diamonds represent $\beta$, as estimated using $Q$ from the SEM analysis, while error bars represent the uncertainty of the mean $\beta$ values and were estimated from propagating $p_{\text{limb}}$ and $Q$ uncertainties.

# References

1. Kim, D. Y., Nishiyama, Y., Wada, M. & Kuga, S. High-yield carbonization of cellulose by sulfuric acid impregnation. *Cellulose* **8**, 29–33 (2001).
2. Frka-Petesic, B., Kamita, G., Guidetti, G. & Vignolini, S. Angular optical response of cellulose nanocrystal films explained by the distortion of the arrested suspension upon drying. *Phys. Rev. Mater.* **3**, 045601 (2019).
3. Frka-Petesic, B., Guidetti, G., Kamita, G. & Vignolini, S. Controlling the Photonic Properties of Cholesteric Cellulose Nanocrystal Films with Magnets. *Adv. Mater.* **29**, 1701469 (2017).
4. Parker, R. M. *et al.* Hierarchical Self-Assembly of Cellulose Nanocrystals in a Confined Geometry. *ACS Nano* **10**, 8443–8449 (2016).
5. Geng, Y., Noh, J. H., Drevensek-Olenik, I., Rupp, R. & Lagerwall, J. Elucidating the fine details of cholesteric liquid crystal shell reflection patterns. *Liq. Cryst.* **44**, 1948–1959 (2017).
6. Noh, J., Liang, H. L., Drevensek-Olenik, I. & Lagerwall, J. P. F. Tuneable multicoloured patterns from photonic cross-communication between cholesteric liquid crystal droplets. *J. Mater. Chem. C* **2**, 806–810 (2014).
7. Chan, C. L. C. *et al.* Visual Appearance of Chiral Nematic Cellulose-Based Photonic Films: Angular and Polarization Independent Color Response with a Twist. *Adv. Mater.* **31**, 1905151 (2019).